\documentclass[a4paper,fleqn,usenatbib]{mnras}

\usepackage[T1]{fontenc}
\usepackage{ae,aecompl}

\usepackage{graphicx}	
\usepackage{amsmath}	
\usepackage{amssymb}	

\newcommand\grad{{\bmath\nabla}}
\newcommand\bmf{{\bmath f}}
\newcommand\bmp{{\bmath p}}
\newcommand\bmq{{\bmath q}}
\newcommand\bmP{{\bmath P}}
\newcommand\bmQ{{\bmath Q}}
\newcommand\calE{\mathcal{E}}
\newcommand\half{{\textstyle\frac{1}{2}}}
\newcommand\rmd{\mathrm{d}}
\newcommand\rme{\mathrm{e}}
\newcommand\rmi{\mathrm{i}}
\newcommand\rms{\mathrm{s}}
\newcommand\f{\frac}
\newcommand\p{\partial}
\newcommand\cst{\mathrm{constant}}

\title[Hamiltonian hydrodynamics of eccentric discs]
{Hamiltonian hydrodynamics of eccentric discs}

\author[Gordon I. Ogilvie and Elliot M.\ Lynch]
{Gordon I. Ogilvie\thanks{E-mail: gio10@cam.ac.uk} and Elliot M.\ Lynch
\\
Department of Applied Mathematics and Theoretical Physics,
University of Cambridge, Centre for Mathematical Sciences,\\
Wilberforce Road, Cambridge CB3 0WA, UK
}

\date{Accepted 2018 December 13. Received 2018 December 12; in original form 2018 October 25}

\pubyear{2018}

\begin{document}
\label{firstpage}
\pagerange{\pageref{firstpage}--\pageref{lastpage}}
\maketitle

\begin{abstract}
  We show that the ideal hydrodynamics of an eccentric astrophysical
  disc can be derived from a variational principle.  The nonlinear
  secular theory describes the slow evolution of a continuous set of
  nested elliptical orbits as a result of the pressure in a thin disc.
  In the artificial but widely considered case of a 2D disc, the
  hydrodynamic Hamiltonian is just the orbit-averaged internal energy
  of the disc, which can be determined from its eccentricity
  distribution using the geometry of the elliptical orbits.  In the
  realistic case of a 3D disc, the Hamiltonian needs to be modified to
  take into account the dynamical vertical structure of the disc.  The
  simplest solutions of the theory are uniformly precessing nonlinear
  eccentric modes, which make the energy stationary subject to the
  angular momentum being fixed.  We present numerical examples of
  nonlinear eccentric modes up to their limiting amplitudes.  Although
  it lacks dissipation, which is important in many astrophysical
  contexts, this formalism allows a simpler theoretical approach to
  the nonlinear dynamics of eccentric discs than that derived from
  stress integrals, and also connects better with established methods
  of celestial mechanics for cases in which the disc interacts
  gravitationally with one or more orbital companion.
\end{abstract}




\section{Introduction}

Eccentric gaseous discs, in which the dominant motion of the fluid
consists of elliptical Keplerian orbits around a central mass, have
numerous applications in astrophysics.  Circumbinary discs around
binary black holes in galactic nuclei, or around young binary stars in
protoplanetary environments, have been found in several numerical
simulations to become eccentric, even if the binary orbit is circular
\citep{2008ApJ...672...83M,2017MNRAS.466.1170M,2017A&A...604A.102T}.
Eccentricity leads to modulation of the accretion rate and affects the
properties of the disc as well as the evolution of the binary
\citep{1996ApJ...467L..77A}.  Similar physics allows the growth of
eccentricity of planets orbiting in gaps or cavities in protoplanetary
discs
\citep{2001A&A...366..263P,2003ApJ...585.1024G,2006A&A...447..369K,2006ApJ...652.1698D,2008MNRAS.384.1242R,2015ApJ...812...94D,2016MNRAS.458.3221T,2017MNRAS.467.4577T,2017MNRAS.464L.114R,2018MNRAS.474.4460R}.
The mechanism is related to those already investigated for
circumstellar discs in `superhump' binary stars
\citep{1988MNRAS.232...35W,1991ApJ...381..259L,1991ApJ...381..268L,2007MNRAS.378..785S,2008A&A...487..671K}
and for planetary rings
\citep{1981ApJ...243.1062G,1983AJ.....88.1560B}.  Circumstellar discs
with companions on sufficiently inclined orbits can undergo
Kozai--Lidov cycles in which the disc becomes highly eccentric
\citep{2014ApJ...792L..33M,2015ApJ...807...75F,2015ApJ...813..105F,2017MNRAS.467.1957Z,2017MNRAS.469.4292L}.
Eccentric discs are also increasingly discussed in connection with
tidal disruption events in which a star on a nearly parabolic orbit
passes too close to the black hole at the centre of a galaxy
\citep[e.g.][]{2014ApJ...783...23G,2016MNRAS.455.2253B,2018MNRAS.tmp.1907C},
or a planetesimal is disrupted around a white dwarf
\citep{2016MNRAS.462.1461M,2018ApJ...852L..22C,2018ApJ...857..135M}.

In many of these applications the eccentricities and their gradients
are not small enough to be well described by the linear theories
discussed by \citet{2002A&A...388..615P}, \citet{2006MNRAS.368.1123G},
\citet{2008MNRAS.388.1372O} and \citet{2016MNRAS.458.3221T}.  A
nonlinear theory of eccentric discs was developed by
\citet{2001MNRAS.325..231O} and rederived from an alternative
viewpoint by \citet{2014MNRAS.445.2621O}.  Evolutionary equations in
one spatial dimension were obtained for the shape and mass
distribution of the disc; however, these equations are complicated
because they contain nonlinear functions of the eccentricity,
eccentricity gradient and twist that must be (pre-)computed
numerically by solving a set of nonlinear ordinary differential
equations around each orbit and evaluating a number of integrals of
stress components combined with geometrical factors.  Perhaps as a
result of this complexity, the nonlinear aspects of the theory have
not been investigated widely in the literature, although inviscid
nonlinear modes have been computed and tested in 2D simulations by
\citet{2016MNRAS.458.3739B}.

The purpose of this paper is to present an alternative, and in some
ways much simpler, approach to the nonlinear dynamics of eccentric
discs.  We focus on the special case in which dissipation can be
neglected; it is then possible to apply a Hamiltonian formulation,
which has attractive mathematical properties and assists in the
physical interpretation of the solutions.  As in the case of the
classical dynamics of systems of particles or rigid bodies,
Hamiltonian methods are useful mainly because they clarify the
mathematical structure of the problem and provide new insights.

It is well known from the linear theory that eccentricity propagates
through gaseous discs predominantly by means of pressure, in the form
of a dispersive wave.  Viscous forces or other dissipative effects
generally contribute a slow diffusion and/or damping of eccentricity.
The ideal fluid theories that we discuss in this paper describe the
nonlinear version of the propagation of eccentricity by means of
pressure, and neglect the weaker effects of viscous forces and other
dissipative effects.  Self-gravity can also make an important
contribution to the propagation of eccentricity, e.g.\ in planetary
rings; it is relatively easy to include within the Hamiltonian
framework but we do not consider it in this paper.

This paper is structured as follows.  We first discuss the geometry of
eccentric discs (Section~\ref{s:geometry}) by defining a set of
canonical variables and analysing the related orbital coordinate
system.  We then derive the evolutionary equations for an eccentric
disc from a Hamiltonian (Section~\ref{s:hamiltonian}) and point out
some of their mathematical properties.  The known linear theory is
verified in Section~\ref{s:linear}.  Section~\ref{s:modes} discusses
special modal solutions of the nonlinear equations and gives numerical
examples.  Conclusions are given in Section~\ref{s:conclusion} and
some mathematical derivations appear in the appendices.

\section{Geometry of eccentric discs}
\label{s:geometry}

Let $(x,y)$ and $(r,\phi)$ be Cartesian and polar coordinates in a
plane, related by
\begin{equation}
  x=r\cos\phi,\qquad
  y=r\sin\phi.
\end{equation}
The polar equation for an elliptical Keplerian orbit of semimajor axis
$a$, eccentricity $e$ and longitude of periapsis $\varpi$ (`curly pi')
is
\begin{equation}
  r=\f{a(1-e^2)}{1+e\cos f},
\end{equation}
where
\begin{equation}
  f=\phi-\varpi
\end{equation}
is the true anomaly.  Equivalently,
\begin{equation}
  r=a(1-e\cos E),
\end{equation}
where $E$ is the eccentric anomaly, which satisfies
\begin{equation}
  \cos f=\f{\cos E-e}{1-e\cos E},\qquad
  \sin f=\f{\sqrt{1-e^2}\sin E}{1-e\cos E}
\label{cosfsinf}
\end{equation}
as well as Kepler's equation
\begin{equation}
  M=E-e\sin E,
\end{equation}
where (\emph{in this Section only}) $M=n(t-\tau)$ is the mean anomaly,
$n=(GM_1/a^3)^{1/2}$ is the mean motion, $M_1$ is the central mass and
$\tau$ is the time of periapsis passage.

For some purposes, such as bringing out the Hamiltonian structure of
the dynamics, it is convenient to use canonical variables such as the
modified Delaunay variables
\begin{align}
&  \Lambda=\sqrt{GM_1a},\qquad
  \lambda=M+\varpi,
\\
&  \Gamma=\Lambda\left(1-\sqrt{1-e^2}\right),\qquad
  \gamma=-\varpi.
\end{align}
These are action--angle variables, with one pair $(\Lambda,\lambda)$
describing the orbital motion and the other pair $(\Gamma,\gamma)$
describing the eccentricity.  More specifically, $\Lambda$ identifies
the orbit and determines its period and energy, while $\lambda$ (the
mean longitude) is an orbital phase that increases linearly in time by
$2\pi$ per orbital period.  Then $\Gamma$ is a positive-definite
measure of the eccentricity of the orbit, being proportional to $e^2$
when $e^2\ll1$, while $\gamma$ determines the orientation of the
ellipse.

A planar eccentric disc involves a continuous set of nested elliptical
orbits.  We can describe the shape of the disc by considering $e$ and
$\varpi$ to be functions of $a$.  The derivatives of these functions
are written as $e_a=\rmd e/\rmd a$ and $\varpi_a=\rmd\varpi/\rmd a$.
Equivalently, when using modified Delaunay variables, we can consider
$\Gamma$ and $\gamma$ to be functions of $\Lambda$, with derivatives
$\Gamma_\Lambda=\rmd\Gamma/\rmd\Lambda$ and
$\gamma_\Lambda=\rmd\gamma/\rmd\Lambda$.  Thus $e_a$ (or
$\Gamma_\Lambda-\Gamma/\Lambda$) is a measure of the eccentricity
gradient, while $\varpi_a$ (or $\gamma_\Lambda$) is a measure of the
twist.

We can use $(\Lambda,\lambda)$ as a canonical orbital coordinate
system covering the disc.  The first coordinate $\Lambda$ labels the
elliptical orbits, while around each orbit $\lambda$ ranges from $0$
to $2\pi$. The Cartesian coordinates $(x,y)$ can be deduced from
$(\Lambda,\lambda)$ as follows.  The value of $\Lambda$ determines $a$
and hence $e$ and $\varpi$ through the functions $e(a)$ and
$\varpi(a)$.  The value of $\lambda$ determines the mean anomaly
$M=\lambda-\varpi$.  Kepler's equation can be solved to find the
eccentric anomaly $E$.  We can then find $f$ from
equations~(\ref{cosfsinf}).  Thus we obtain $(r,\phi)$ and $(x,y)$.

In previous work \citep{2001MNRAS.325..231O,2014MNRAS.445.2621O} we
have instead used $(\lambda,\phi)$ as an orbital coordinate system,
where (\emph{in this paragraph only}) $\lambda=a(1-e^2)$ is the semilatus
rectum.  The geometry of eccentric discs is slightly easier to analyse
when $e$ and $\varpi$ are considered to be functions of $\lambda$
rather than $a$.  Also $\lambda$ is directly related to the orbital
angular momentum, which is often more important than the orbital
energy in the context of accretion discs.  However, for the
non-dissipative discs described by Hamiltonian hydrodynamics, it is
more convenient to label the orbits using $a$, which is directly
related to the orbital period and energy, and is a material invariant
in the Hamiltonian theory.  This approach also works better when
combined with secular gravitational interactions, which similarly
leave $a$ invariant.

By differentiating the above relations, we find that the Jacobian of
the canonical orbital coordinate system $(\Lambda,\lambda)$ is
\begin{equation}
  J=\f{\p(x,y)}{\p(\Lambda,\lambda)}=J^\circ(a)j(E),
\end{equation}
where
\begin{equation}
  J^\circ=\f{2}{n}
\end{equation}
is the Jacobian for a circular disc and
\begin{align}
  j&=1-\f{\Gamma}{2\Lambda}-\f{\Gamma_\Lambda}{2}+\f{(\Gamma-\Lambda\Gamma_\Lambda)\cos E}{2\sqrt{\Gamma(2\Lambda-\Gamma)}}\nonumber\\
  &\qquad+\f{\sqrt{\Gamma(2\Lambda-\Gamma)}\gamma_\Lambda\sin E}{2}\nonumber\\
  &=\f{1-e(e+ae_a)}{\sqrt{1-e^2}}-\f{ae_a\cos E}{\sqrt{1-e^2}}-ae\varpi_a\sin E
\end{align}
is dimensionless and purely related to the elliptical geometry of the
disc.  It can be verified that $J$ remains positive throughout the
orbit provided that $(e+ae_a)^2+(ae\varpi_a)^2<1$.  This condition was
given by \citet{2001AJ....122.2257S} and is exactly equivalent to the
criterion $(e-\lambda e_\lambda)^2+(\lambda e\varpi_\lambda)^2<1$
given by \citet{2001MNRAS.325..231O} in terms of the semilatus
rectum. It is the condition for the eccentricity, eccentricity
gradient and twist to be sufficiently small that neighbouring orbits
do not intersect.  Of course $e^2<1$ is also required for the orbits
to be closed.

In the special case of an untwisted eccentric disc, for which
$\varpi_a=0$, the condition for marginal orbital intersection,
$e+ae_a=\pm1$, is equivalent to
\begin{equation}
  \f{\rmd}{\rmd a}(a\mp ae)=0,
\end{equation}
i.e.\ vanishing derivative of either periapsis or apoapsis distance
with $a$.  The orbits intersect when either the periapsis or apoapsis
distance \emph{decreases} with $a$.

The orbital coordinate system is not orthogonal, but fortunately the
Hamiltonian approach does not require any tensor calculus.  The
contravariant components of the orbital velocity are $u^\Lambda=0$ and
$u^\lambda=n=(GM_1)^2/\Lambda^3$, which depends only on $\Lambda$.

Assuming that the surface density $\Sigma$ is stationary in an
inertial frame on the orbital timescale, the equation of mass
conservation implies
\begin{equation}
  \f{\p}{\p\lambda}(J\Sigma n)=0.
\end{equation}
Therefore $J\Sigma$ is independent of $\lambda$. Since the mass of the
disc is given by the integral
\begin{equation}
  \int\Sigma\,\rmd A=\iint\Sigma\,J\,\rmd\Lambda\,\rmd\lambda=2\pi\int J\Sigma\,\rmd\Lambda,
\end{equation}
we see that
\begin{equation}
  2\pi J\Sigma=M_\Lambda,
\label{sigma}
\end{equation}
where $M(\Lambda)$ is the mass of the disc contained within the orbit
labelled by $\Lambda$, and $M_\Lambda=\rmd M/\rmd\Lambda$ is the
one-dimensional mass density with respect to $\Lambda$.  In the
Hamiltonian theory $M(\Lambda)$ does not depend on time because
$\Lambda$ is a material invariant.

\section{Hamiltonian evolution}
\label{s:hamiltonian}

The secular theory of celestial mechanics describes the slow evolution
of the orbits of a number of bodies around a central mass under the
assumption that the mutual gravitational interactions of the bodies
are relatively weak and do not involve mean-motion resonances.  The
secular Hamiltonian is just the mutual gravitational energy of the
interacting bodies, averaged over their orbital motion.  Since the
Hamiltonian does not depend on the orbital longitude of any of the
bodies, each orbit preserves its semimajor axis (and energy).  Angular
momentum is exchanged between the orbits, resulting in precession and
oscillations of eccentricity (as well as inclination, if a non-planar
system is considered).

The equivalent situation for a non-self-gravitating, ideal fluid disc
is one in which forces due to pressure gradients are small compared to
the gravity of the central mass.  This condition is satisfied in a
thin disc, provided that the lengthscale associated with the
eccentricity gradient or twist is long compared to the thickness of
the disc.  The disc can then be treated as a continuous set of nested
elliptical orbits that evolve slowly due to pressure.  We will see
that the Hamiltonian describing this evolution is just equal to the
internal energy of the system, in the artificial case of a 2D disc,
while in a 3D disc there is a modification due to the dynamical
vertical structure.  Again, the orbital energy, and therefore the
semimajor axis, is a material invariant; this is true because no
energy is dissipated, nor is any energy transferred to neighbouring
orbits, because the energy flux density associated with a fluid
pressure is always parallel to the velocity.  Therefore the secular
hydrodynamics of an eccentric disc amounts to determining how the
shape of the disc evolves in time. We can obtain equations for the
time-derivatives of the functions $e(a,t)$ and $\varpi(a,t)$,
describing this evolution. In fact we will do this first in the
equivalent canonical variables, obtaining evolutionary equations for
$\Gamma(\Lambda,t)$ and $\gamma(\Lambda,t)$.

If the secular Hamiltonian of a fluid disc is written as
\begin{equation}
  H=\int H_\Lambda\,\rmd\Lambda,
\end{equation}
with the Hamiltonian density $H_\Lambda$ being expressible a function
of $(\Gamma,\Gamma_\Lambda,\gamma,\gamma_\Lambda,\Lambda)$, then the
canonical form of the evolutionary equations is
\begin{align}
&  M_\Lambda\f{\p\Gamma}{\p t}=-\f{\delta H}{\delta\gamma},
\label{dGammadt}\\
&  M_\Lambda\f{\p\gamma}{\p t}=\f{\delta H}{\delta\Gamma},
\end{align}
where the $\delta$ notation represents a functional derivative, i.e.
\begin{align}
&  \f{\delta H}{\delta\Gamma}=\f{\p H_\Lambda}{\p\Gamma}-\f{\p}{\p\Lambda}\left(\f{\p H_\Lambda}{\p\Gamma_\Lambda}\right),
\\
&  \f{\delta H}{\delta\gamma}=\f{\p H_\Lambda}{\p\gamma}-\f{\p}{\p\Lambda}\left(\f{\p H_\Lambda}{\p\gamma_\Lambda}\right),
\end{align}
as appears in the Euler--Lagrange equation of variational calculus.
Indeed 
the Hamiltonian $H$ is a functional of the two functions
$\Gamma(\Lambda)$ and $\gamma(\Lambda)$ that describe the shape of the
eccentric disc at any instant of time.

The rotational invariance of the system means that $H_\Lambda$ does
not depend explicitly on $\gamma$, as we shall see in detail
below. Therefore equation~(\ref{dGammadt}) simplifies to
\begin{equation}
  M_\Lambda\f{\p\Gamma}{\p t}=\f{\p}{\p\Lambda}\left(\f{\p H_\Lambda}{\p\gamma_\Lambda}\right),
\end{equation}
which can be written in the conservative form
\begin{equation}
  \f{\p}{\p t}(M_\Lambda\Gamma)+\f{\p}{\p\Lambda}\left(-\f{\p H_\Lambda}{\p\gamma_\Lambda}\right)=0.
\end{equation}
With suitable boundary conditions, such that the flux
$-\p H_\Lambda/\p\gamma_\Lambda$ vanishes at the boundaries of the
disc, the total angular-momentum deficit (AMD)
\begin{equation}
  C=\int M_\Lambda\Gamma\,\rmd\Lambda=\int\Gamma\,\rmd M
\label{amd}
\end{equation}
is conserved. The AMD is a positive-definite measure of the
eccentricity of a system, being proportional to $e^2$ when $e^2\ll1$,
and is widely used in modern celestial mechanics
\citep{1997A&A...317L..75L}.  It is the difference between the angular
momentum of the set of elliptical orbits and that of a set of circular
orbits with the same semimajor axes.  The AMD is conserved because the
angular momentum is conserved and $a$ is a material invariant.  The
transport of angular momentum, or of AMD, is related to the twisting
of the eccentric disc, because the flux
$-\p H_\Lambda/\p\gamma_\Lambda$ is found to vanish in the untwisted
case $\varpi_a=0$.

The total Hamiltonian $H$ is also conserved.  Since the Hamiltonian
density does not depend explicitly on time, it evolves according to
\begin{equation}
  \f{\p H_\Lambda}{\p t}=\f{\p H_\Lambda}{\p\Gamma}\f{\p\Gamma}{\p t}+\f{\p H_\Lambda}{\p\gamma}\f{\p\gamma}{\p t}+\f{\p H_\Lambda}{\p\Gamma_\Lambda}\f{\p\Gamma_\Lambda}{\p t}+\f{\p H_\Lambda}{\p\gamma_\Lambda}\f{\p\gamma_\Lambda}{\p t}.
\end{equation}
The notation here deserves some comment.  On the left-hand side of
this equation we are considering $H_\Lambda$ as a function of
$(\Lambda,t)$, describing its spatial distribution with the disc and
its evolution in time.  On the right-hand side we are considering
$H_\Lambda$ as a function of the geometrical--dynamical parameters
$(\Gamma,\Gamma_\Lambda,\gamma,\gamma_\Lambda,\Lambda)$, each of which
is itself a function of $(\Lambda,t)$.  This equation can be written
in the conservative form
\begin{align}
  \f{\p H_\Lambda}{\p t}+\f{\p}{\p\Lambda}\left(-\f{\p H_\Lambda}{\p\Gamma_\Lambda}\f{\p\Gamma}{\p t}-\f{\p H_\Lambda}{\p\gamma_\Lambda}\f{\p\gamma}{\p t}\right)&=\f{\delta H}{\delta\Gamma}\f{\p\Gamma}{\p t}+\f{\delta H}{\delta\gamma}\f{\p\gamma}{\p t}\nonumber\\
  &=0.
\end{align}
Therefore both $C$ and $H$ are conserved if
$\p H_\Lambda/\p\Gamma_\Lambda$ and $\p H_\Lambda/\p\gamma_\Lambda$
vanish at the boundaries.  These two conservation laws are instances
of Noether's Theorem relating conservation laws to continuous
symmetries.

The equivalent evolutionary equations for the more familiar but
non-canonical variables $e(a,t)$ and $\varpi(a,t)$ are
\begin{align}
&  M_a\f{\p e}{\p t}=\f{\sqrt{1-e^2}}{na^2e}\f{\delta H}{\delta\varpi},
\label{edot}\\
&  M_a\f{\p\varpi}{\p t}=-\f{\sqrt{1-e^2}}{na^2e}\f{\delta H}{\delta e},
\label{pidot}
\end{align}
where $M_a=\rmd M/\rmd a$ is the one-dimensional mass density with
respect to $a$.  Note that $M_a/M_\Lambda=\rmd\Lambda/\rmd a=na/2$.
In this case the relevant functional derivatives are
\begin{align}
&  \f{\delta H}{\delta e}=\f{\p H_a}{\p e}-\f{\p}{\p a}\left(\f{\p H_a}{\p e_a}\right),
\\
&  \f{\delta H}{\delta\varpi}=\f{\p H_a}{\p\varpi}-\f{\p}{\p a}\left(\f{\p H_a}{\p\varpi_a}\right).
\end{align}
From the fact that $H_a$ does not depend explicitly on $\varpi$,
it is similarly possible to deduce the conservation of the AMD in the
form
\begin{equation}
  C=\int M_ana^2\left(1-\sqrt{1-e^2}\right)\,\rmd a,
\end{equation}
which is equivalent to equation~(\ref{amd}).

We present a detailed argument in Appendix~\ref{s:appendixa} for why
the secular hydrodynamics a non-self-gravitating, ideal fluid disc can
be derived from Hamilton's equations in the above form.  In the
artificial case of a 2D disc, the Hamiltonian density is
\begin{equation}
  H_\Lambda^{\mathrm{(2D)}}=M_\Lambda\langle\varepsilon\rangle,\qquad\hbox{or}\qquad
  H_a^{\mathrm{(2D)}}=M_a\langle\varepsilon\rangle,
\end{equation}
where $\varepsilon$ is the specific internal energy and the angle
brackets denote an orbital average.  In Appendix~\ref{s:appendixb} we
show that this can be written as
\begin{equation}
  H_\Lambda^{\mathrm{(2D)}}=H_\Lambda^\circ F^\mathrm{(2D)},\qquad\hbox{or}\qquad
  H_a^{\mathrm{(2D)}}=H_a^\circ F^\mathrm{(2D)},
\end{equation}
where, for a perfect gas of adiabatic index\footnote{There is an
  unfortunate notational clash between the adiabatic index and one of
  the modified Delaunay variables.  The context should make it clear
  which meaning $\gamma$ has in any equation.} $\gamma$,
\begin{equation}
  F^\mathrm{(2D)}=\f{1}{\gamma-1}\left\langle j^{-(\gamma-1)}\right\rangle
\end{equation}
is the dimensionless, `geometric' part of the Hamiltonian density,
depending only on $e$, $ae_a$, $ae\varpi_a$ and $\gamma$, while
\begin{equation}
  H_\Lambda^\circ=\f{4\pi P^\circ}{n},\qquad\hbox{or}\qquad
  H_a^\circ=2\pi aP^\circ,
\end{equation}
is a fixed function of $\Lambda$ (or $a$), $P^\circ$ being
the (2D, or vertically integrated) pressure of a circular disc with
the same distributions of mass and entropy.  In the isothermal limit
$\gamma=1$ we have instead
\begin{equation}
  F^\mathrm{(2D)}=-\left\langle\ln j\right\rangle.
\end{equation}
Explicit expressions for $F^\mathrm{(2D)}$ are derived in
Appendix~\ref{s:appendixc}; these involve only elementary functions if
$\gamma=1$ or~$2$ and involve Legendre functions otherwise.

In the realistic case of a 3D disc, the Hamiltonian is modified
because of the dynamical vertical structure of the disc.  The
Hamiltonian density is instead
\begin{equation}
  H_\Lambda=\f{1}{2}(\gamma+1)M_\Lambda\langle\bar\varepsilon\rangle,
\end{equation}
where $\bar\varepsilon$ is the (mass-weighted) vertically averaged
specific internal energy and $\langle\bar\varepsilon\rangle$ is its
orbital average.  In this case
\begin{equation}
  H_\Lambda^{\mathrm{(3D)}}=H_\Lambda^\circ F^\mathrm{(3D)},
\end{equation}
with
\begin{equation}
  F^\mathrm{(3D)}=\f{(\gamma+1)}{2(\gamma-1)}\left\langle(jh)^{-(\gamma-1)}\right\rangle,
\end{equation}
where $h(E)$ describes the variation of the dimensionless vertical
scaleheight around the orbit and must in general be obtained as the
solution of the second-order ordinary differential equation
(ODE)~(\ref{ode}).  Again, the geometric part $F^\mathrm{(3D)}$
depends only on $e$, $ae_a$, $ae\varpi_a$ and $\gamma$.  In the
isothermal limit $\gamma=1$ we have instead
\begin{equation}
  F^\mathrm{(3D)}=-\left\langle\ln(jh)\right\rangle.
\end{equation}

\begin{figure*}
\includegraphics[width=1.1\columnwidth]{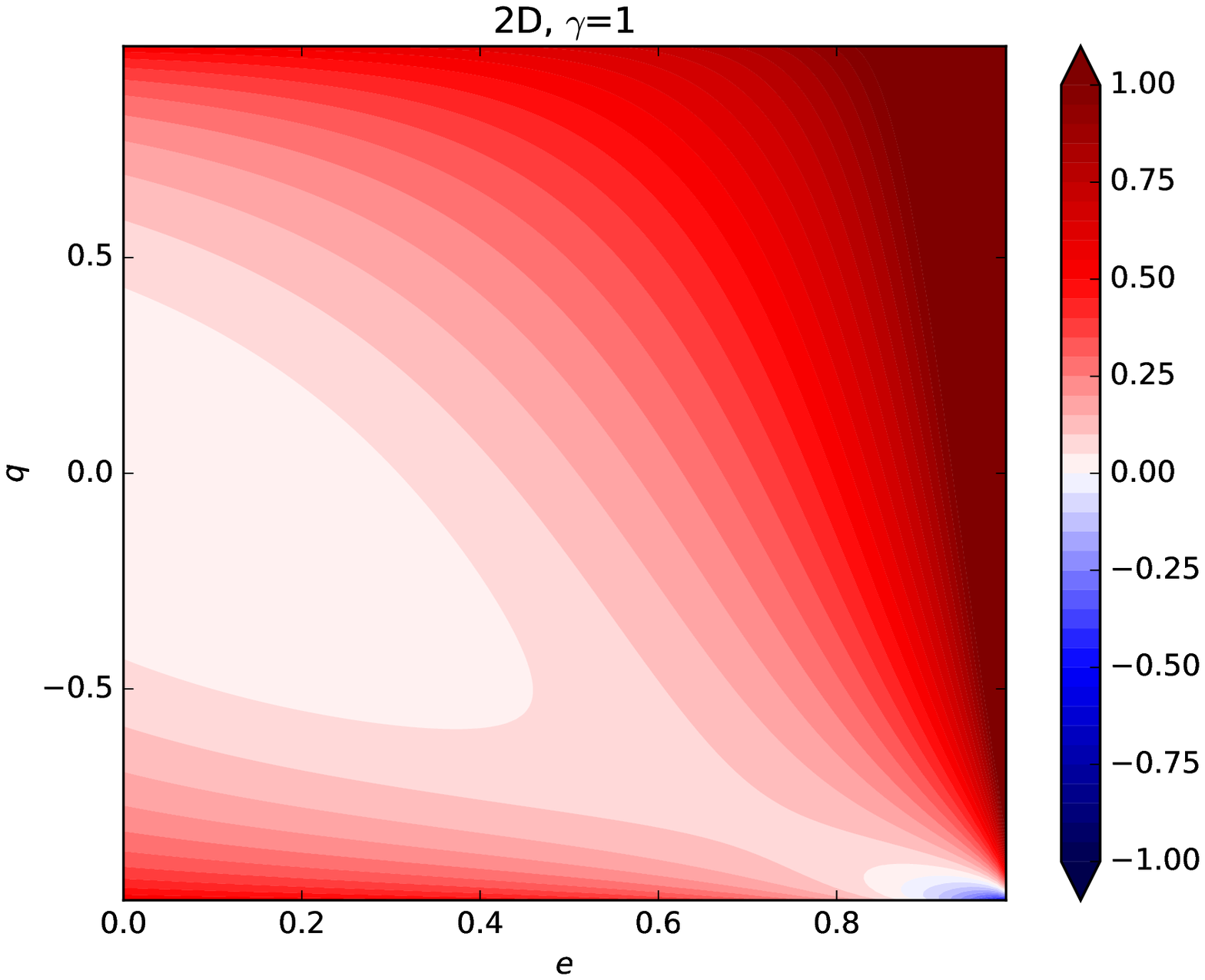}
\hskip-1cm
\includegraphics[width=1.1\columnwidth]{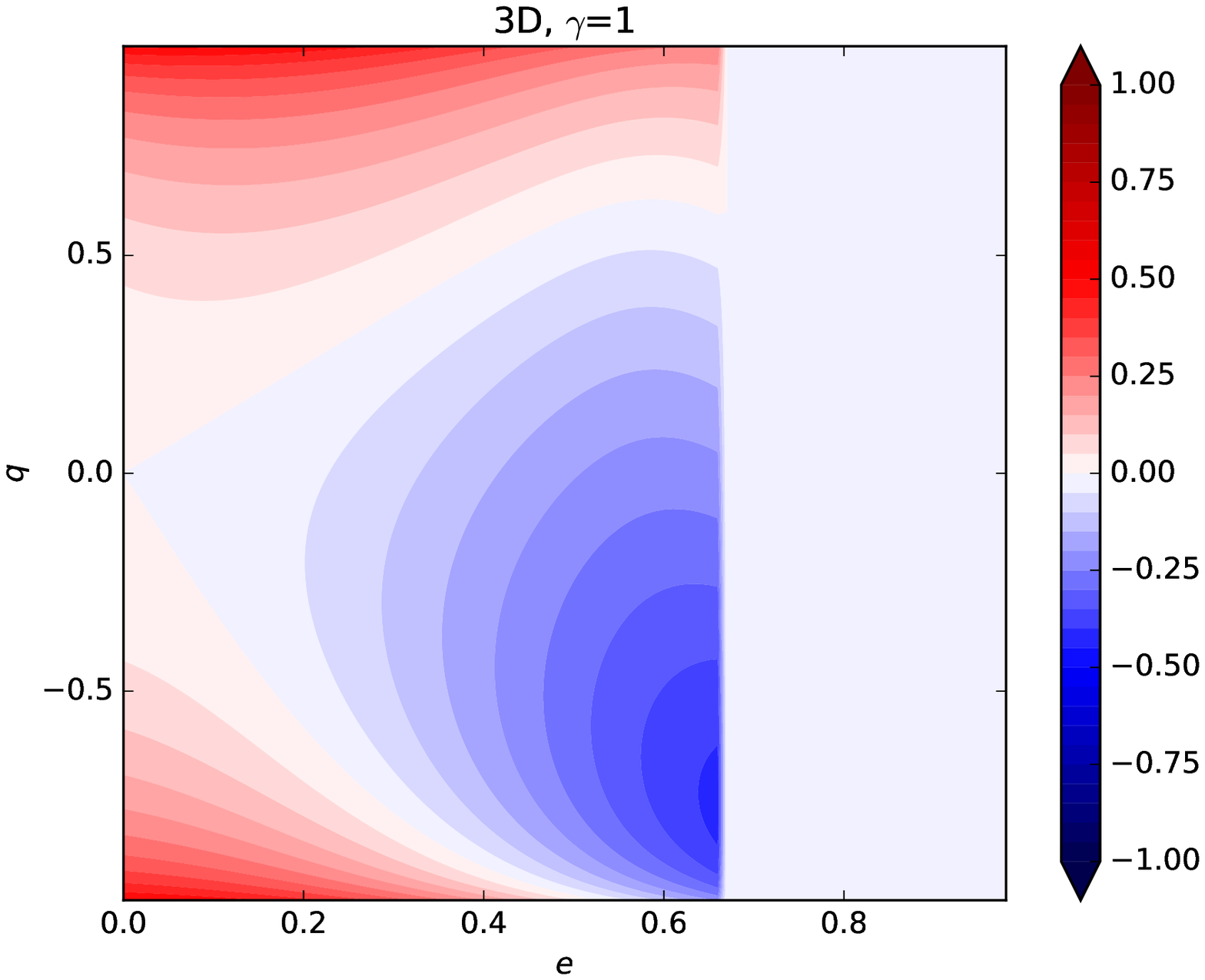}
\hskip1.1\columnwidth
\\
\includegraphics[width=1.1\columnwidth]{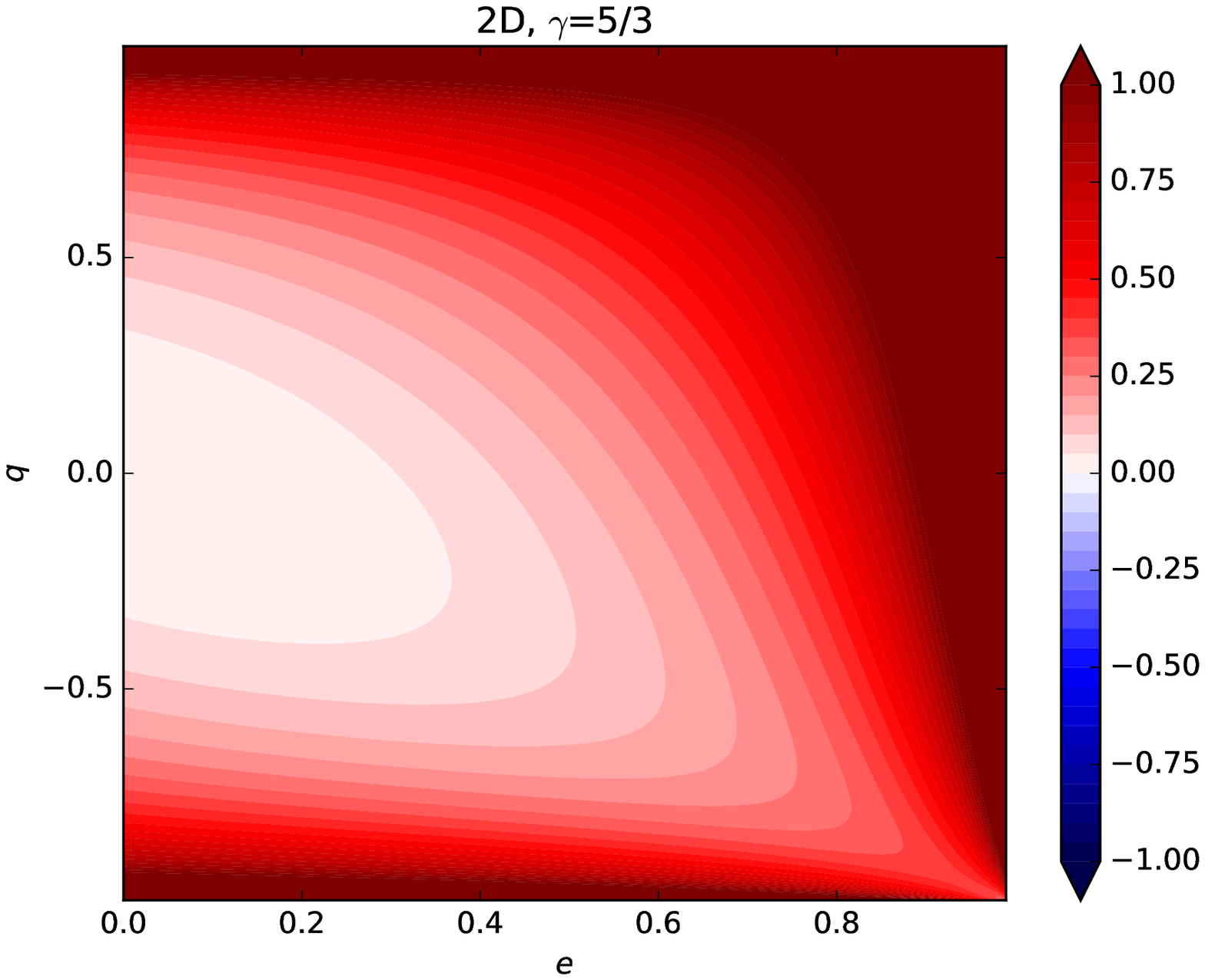}
\hskip-1cm
\includegraphics[width=1.1\columnwidth]{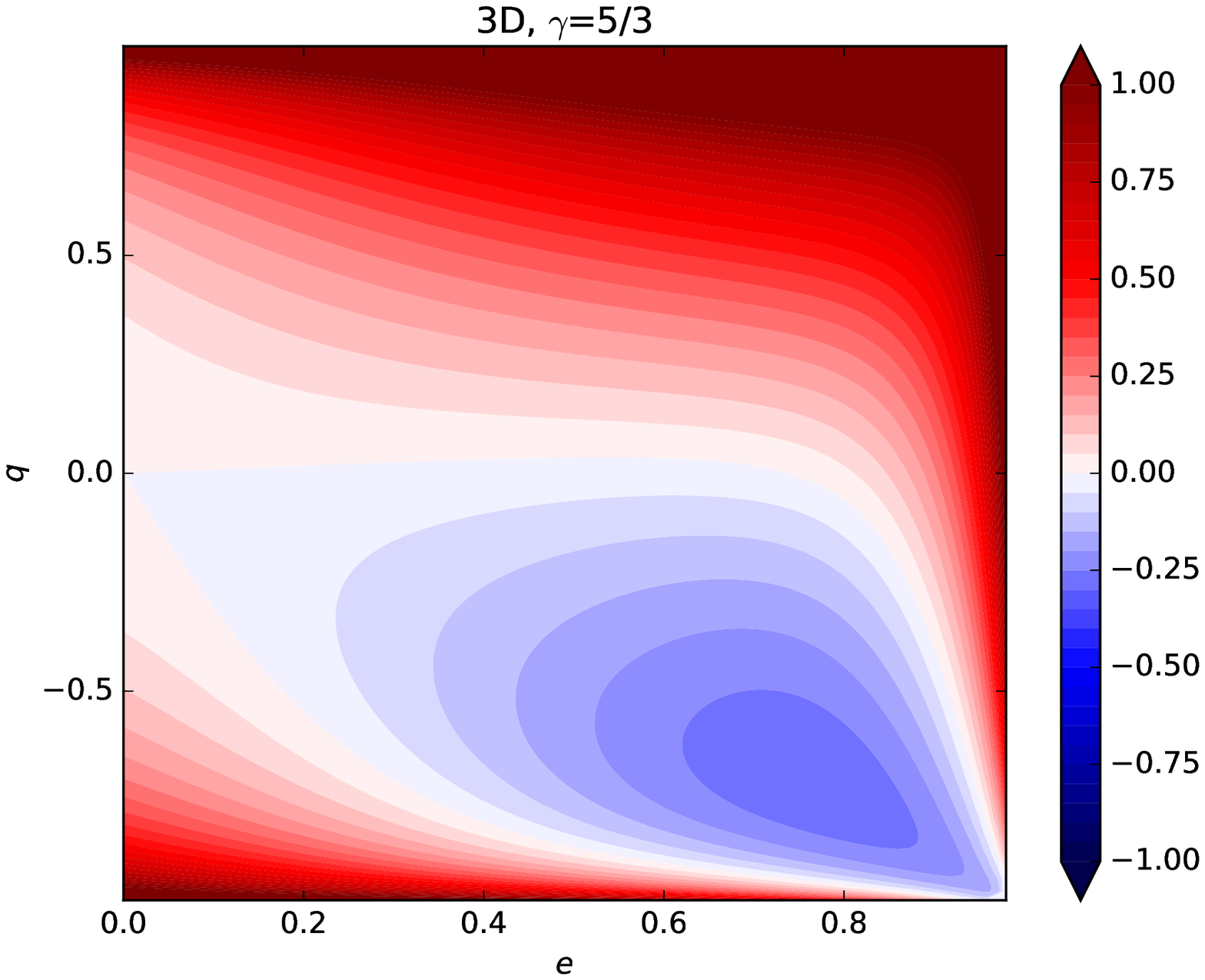}
\caption{Dimensionless geometric part of the Hamiltonian density for
  untwisted eccentric discs, as a function of the eccentricity $e$ and
  the (signed) nonlinearity parameter $q=ae_a/[1-e(e+ae_a)]$.  Top
  left: 2D case for $\gamma=1$.  Top right: 3D case for $\gamma=1$.
  (Values are not plotted for $e>0.66$ because the solution involves
  extreme compression at periapsis.)  Bottom left: 2D case for
  $\gamma=5/3$, with the additive constant term removed.  Bottom
  right: 3D case for $\gamma=5/3$, with the additive constant term
  removed.}
\label{f:geometric}
\end{figure*}

In Fig.~\ref{f:geometric} the geometric parts of the 2D and 3D
Hamiltonians are compared in the case of untwisted discs. The four
panels show how much the dynamics of an eccentric disc is affected by
its thermodynamic behaviour and, in particular, by the dynamical
vertical structure of a 3D disc.

\section{Verification of linear theory}
\label{s:linear}

When $e$, $ae_a$ and $ae\varpi_a$ are much less than unity, the
geometric part of the Hamiltonian density for a 2D disc can be
approximated as the quadratic function
\begin{equation}
  F^\mathrm{(2D)}\approx\f{1}{\gamma-1}+\f{1}{2}e(e+ae_a)+\f{1}{4}\gamma[(ae_a)^2+(ae\varpi_a)^2].
\label{f2dlinear}
\end{equation}
The same expression holds in the isothermal case $\gamma=1$ except
that the unimportant constant term is missing.  Using this
approximation for the Hamiltonian in equations (\ref{edot}) and
(\ref{pidot}) leads, after some algebra, to the linearized
evolutionary equation
\begin{equation}
  2\Sigma^\circ na^3\f{\p\calE}{\p t}=\f{\p}{\p a}\left(\rmi\gamma P^\circ a^3\f{\p\calE}{\p a}\right)+\rmi\f{\rmd P^\circ}{\rmd a}a^2\calE,
\end{equation}
where $\calE=e\,\rme^{\rmi\varpi}$ is the complex eccentricity, while
$\Sigma^\circ=M_a/2\pi a$ is the surface density of a circular disc
with the same mass distribution.  This equation is exactly equivalent
to equation (21) of \citet{2006MNRAS.368.1123G} for a 2D adiabatic
disc.

In the same linear regime for a 3D disc, an expansion of the ODE
(\ref{ode}) leads to the approximations
\begin{equation}
  h\approx1-\f{3}{\gamma}e\cos E+\f{(\gamma-1)}{\gamma}(ae_a\cos E+ae\varpi_a\sin E),
\end{equation}
\begin{multline}
  F^\mathrm{(3D)}\approx\f{\gamma+1}{2(\gamma-1)}+\f{1}{4\gamma}\Big\{(5\gamma-9\gamma)e^2+2(4\gamma-3)eae_a\\
  +(2\gamma-1)[(ae_a)^2+(ae\varpi_a)^2]\Big\}.
\end{multline}
This is significantly different from the 2D version and leads instead
to the linearized evolutionary equation
\begin{multline}
  -2\rmi\Sigma^\circ na^3\f{\p\calE}{\p t}=\f{\p}{\p a}\left[\left(2-\f{1}{\gamma}\right)P^\circ a^3\f{\p\calE}{\p a}\right]\\
  +\left(4-\f{3}{\gamma}\right)\f{\rmd P^\circ}{\rmd a}a^2\calE+3\left(1+\f{1}{\gamma}\right)P^\circ a\calE,
\end{multline}
which is exactly equivalent to equation~(2) of
\citet{2016MNRAS.458.3221T} for a 3D adiabatic disc.  The last term,
in particular, is a prograde contribution to the precession of the
disc due to pressure.

The difference between the 2D and 3D theories is most clearly seen in
the isothermal case $\gamma=1$, for which
\begin{equation}
  F^\mathrm{(3D)}=-\langle\ln j\rangle-\langle\ln h\rangle=F^\mathrm{(2D)}-\langle\ln h\rangle.
\end{equation}
Furthermore, in this case the eccentricity gradient and twist do not
affect the vertical oscillation, so $F^\mathrm{(3D)}$ separates
cleanly into horizontal and vertical parts, where the vertical part
depends only on $e$ and has the expansion
\begin{equation}
  -\langle\ln h\rangle=-\f{3}{2}e^2+\f{3}{8}e^4+\f{25}{56}e^6+O(e^8).
\end{equation}
It is a decreasing function of $e$ and so contributes to prograde
precession, according to equation~(\ref{pidot}).

\section{Nonlinear eccentric modes}
\label{s:modes}

An eccentric mode is a special solution in which the eccentricity
distribution $e(a)$ is independent of time and the disc is untwisted
and precesses uniformly at angular frequency $\omega$, such that
\begin{equation}
  \varpi=\omega t+\cst.
\end{equation}
The Hamiltonians we are interested in have the property that
$\p H_a/\p\varpi_a=0$ in an untwisted disc with $\varpi_a=0$.
Therefore equation~(\ref{edot}) is automatically satisfied in an
eccentric mode, while equation~(\ref{pidot}) reduces to
\begin{equation}
  M_a\omega=-\f{\sqrt{1-e^2}}{na^2e}\f{\delta H}{\delta e}.
\label{modal}
\end{equation}
This is a nonlinear ODE for $e(a)$ in which the precession rate
$\omega$ appears as an eigenvalue.  In evaluating the right-hand side
of this equation, the twist $\varpi_a$ can be set to zero.

The modal equation~(\ref{modal}) can also be written in the form
\begin{equation}
  \omega\f{\delta L}{\delta e}=\f{\delta H}{\delta e},
\label{modal2}
\end{equation}
where
\begin{equation}
  L=\int M_ana^2\sqrt{1-e^2}\,\rmd a
\end{equation}
is the angular momentum. Note that $L+C=\int na^2\,\rmd M=L^\circ$ is
a constant property of the disc, being the total angular momentum of a
circular disc with the same mass distribution.

The solutions of the modal equation generally involve oscillations of
the function $e(a)$, similar to waves on a string.  It is conventional
in celestial mechanics to regard $e$ as a non-negative quantity and
$\varpi$ as being defined modulo $2\pi$ wherever $e\ne0$. In the case
of an untwisted eccentric mode, however, where $e(a)$ may have zeros
on particular orbits, it is more convenient to allow $e$ to take both
positive and negative values such that $e_a$ and $\varpi$ vary
continuously.  This alternative viewpoint is possible because of the
invariance of the equations (and the geometry) when $(e,\varpi)$ are
replaced with $(-e,\varpi+\pi)$.

Equation~(\ref{modal2}) is the Euler--Lagrange equation for the
nonlinear variational problem in which we seek stationary values of
$H$ subject to the constraint $L=\cst$ (or, equivalently, $C=\cst$).
The boundary conditions should be such that either the eccentricity
vanishes (a rigid, circular boundary) or $\p H_a/\p e_a=0$ (a free,
vacuum boundary).  The precession frequency $\omega$ appears here as a
Lagrange multiplier.  This equation implies that, when the shape of
the disc undergoes an infinitesimal change from a modal solution, the
infinitesimal changes in $H$ and $L$ are related by
$\delta H=\omega\,\delta L$. We expect there to exist multiple
branches of modal solutions, along each of which the shape of the mode
varies continuously, as do the values of $H$, $L$ and $\omega$. Some
or all of these branches will reach the limit of a circular disc in
which $H\to H^\circ$ and $L\to L^\circ$, while $\omega$ tends to a
constant value which is the precession rate of the mode in linear
theory. Along each branch we expect $H$ and $L$ to be related by
$\rmd H/\rmd L=\omega$.

When the Hamiltonian density for an untwisted disc is written as
$H_a=H_a^\circ F$, where $H_a^\circ$ depends only on $a$, and $F$ is a
dimensionless function of $e$ and $f=e+ae_a$, the modal equation can
be written explicitly as
\begin{multline}
  -\f{\omega M_a}{H_a^\circ}\f{na^2e}{\sqrt{1-e^2}}=\f{\p F}{\p e}-ae_a\f{\p^2F}{\p e\,\p f}-a(2e_a+ae_{aa})\f{\p^2F}{\p f^2}\\
  -\f{\rmd\ln(H_a^\circ)}{\rmd\ln a}\f{\p F}{\p f}.
\label{modal3}
\end{multline}

\begin{figure*}
\includegraphics[width=\columnwidth]{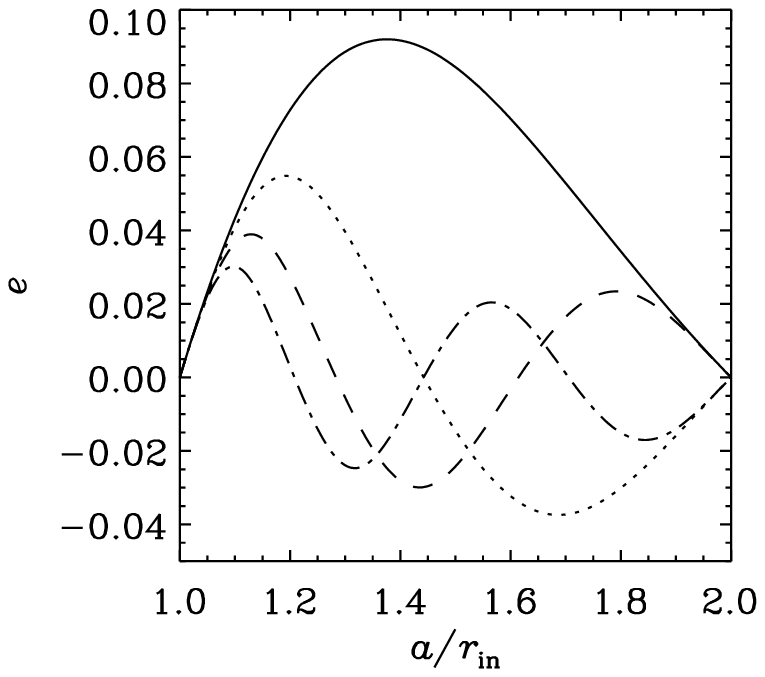}
\includegraphics[width=\columnwidth]{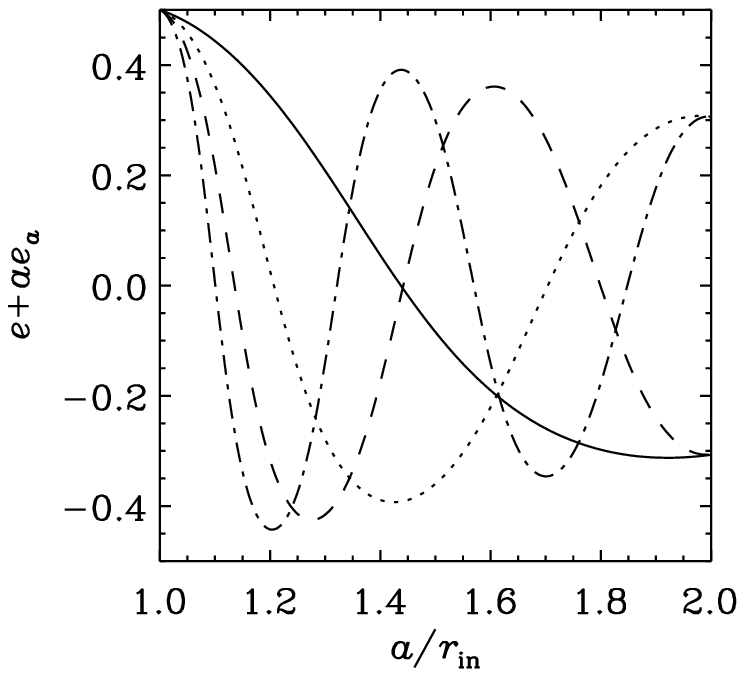}
\\
\includegraphics[width=\columnwidth]{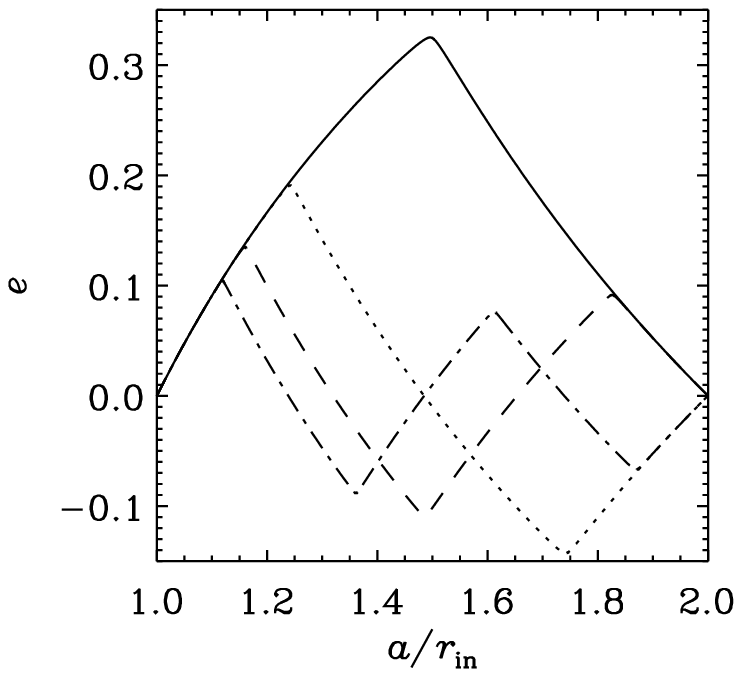}
\includegraphics[width=\columnwidth]{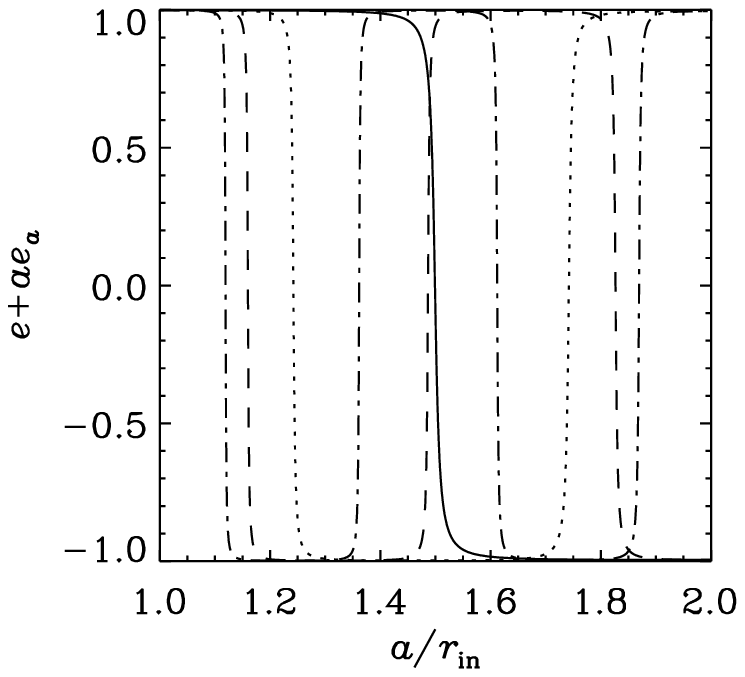}
\caption{Shapes of nonlinear eccentric modes for the simple model
  problem considered by \citet{2016MNRAS.458.3739B} in which a 2D
  isothermal disc is contained within rigid circular boundaries with a
  radius ratio of~$2$.  The upper two panels show the profiles of $e$
  and $f=e+ae_a$ for the first four modes with moderately large
  amplitudes such that $ae_a=0.5$ at the inner boundary.  The lower
  panels show the same modes with nearly maximal amplitudes such that
  $ae_a=0.999$ at the inner boundary.  Only the lowest-order mode was
  computed by \citet{2016MNRAS.458.3739B}.}
\label{f:mode_shape_isothermal_rigid}
\end{figure*}

\begin{figure*}
\includegraphics[width=\columnwidth]{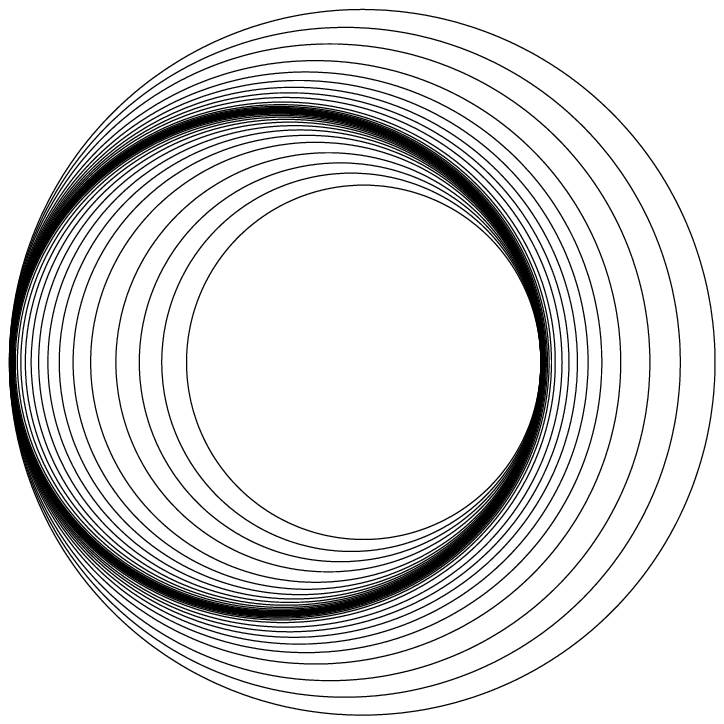}
\includegraphics[width=\columnwidth]{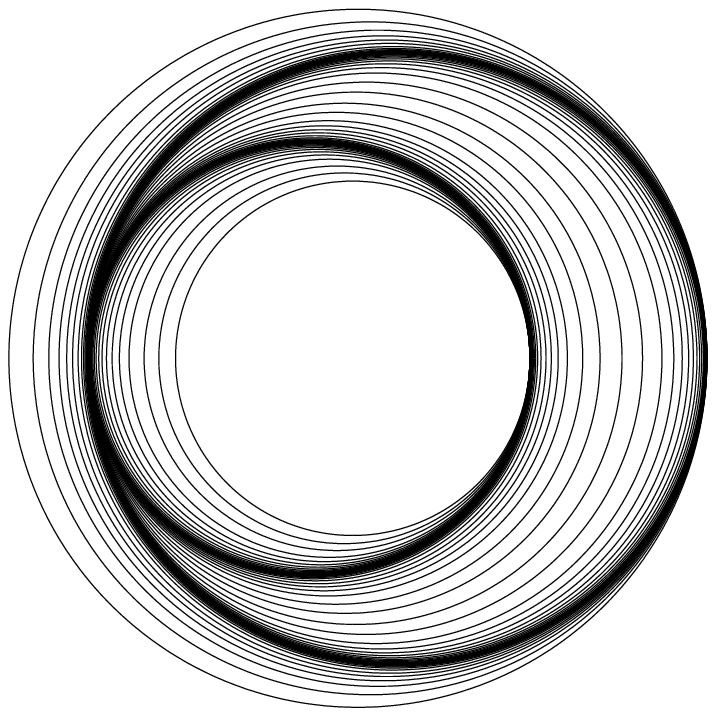}
\\
\includegraphics[width=\columnwidth]{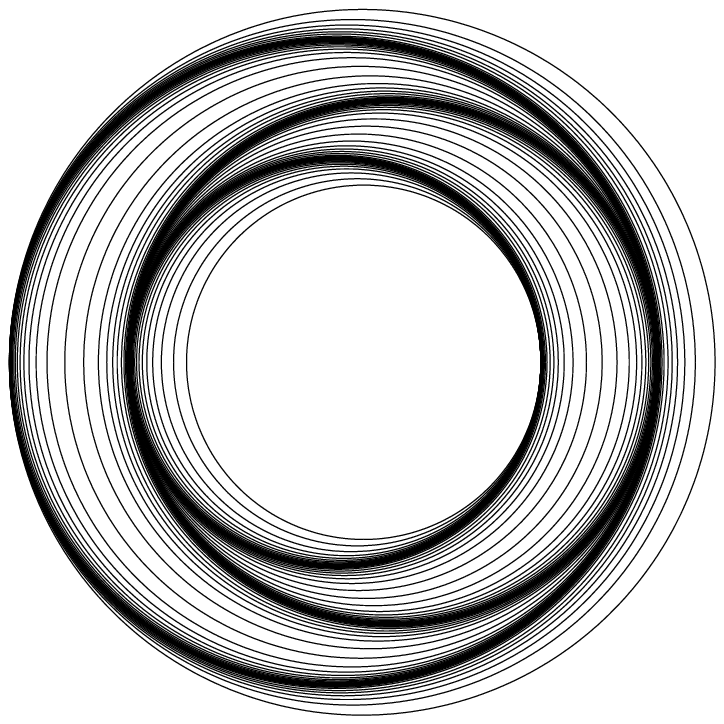}
\includegraphics[width=\columnwidth]{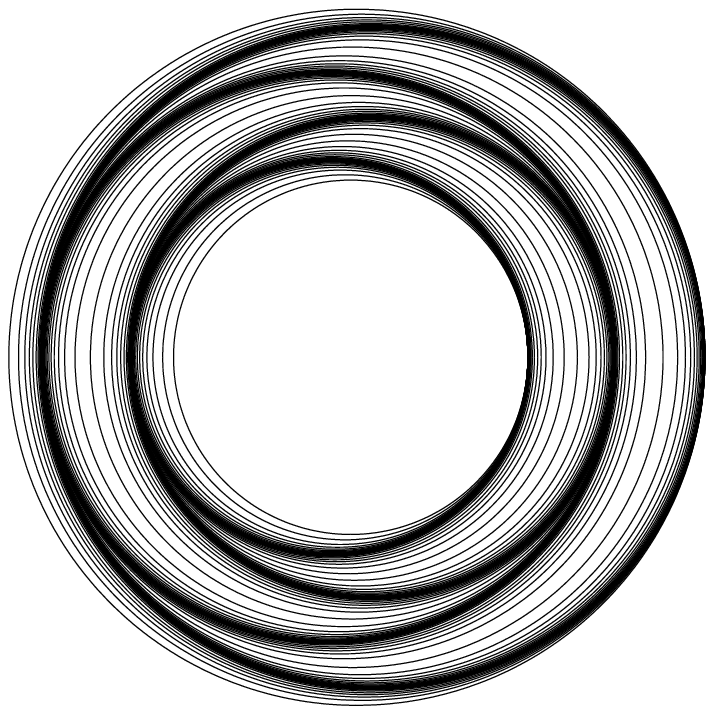}
\caption{Shapes of nonlinear eccentric modes for the same problem as
  in Fig.~\ref{f:mode_shape_isothermal_rigid}, for the first four
  modes with nearly maximal amplitudes such that $ae_a=0.999$ at the
  inner boundary.}
\label{f:mode_orbits_isothermal_rigid}
\end{figure*}

We illustrate this theory by considering the case of a 2D disc.  We
first solve the simple model problem studied by
\citet{2016MNRAS.458.3739B} in which a 2D isothermal disc is contained
within rigid circular boundaries with a radius ratio of~$2$.  The mass
distribution is such that $M_a\propto a$, which corresponds to a
uniform surface density in the limit of a circular disc.  We solve the
second-order nonlinear ODE~(\ref{modal}) as an eigenvalue problem for
the mode frequency $\omega$, using a shooting method.  The boundary
conditions are $e=0$ at $a=r_\mathrm{in}$ and at
$a=r_\mathrm{out}=2r_\mathrm{in}$.  We obtain a sequence of modes with
increasing numbers of nodes in the eigenfunction $e(a)$
(Fig.~\ref{f:mode_shape_isothermal_rigid}).  The amplitude of the mode
can be varied by choosing the value (positive, without loss of
generality) of $ae_a$ at the inner boundary.  As this value approaches
its maximum possible value of~$1$, the modes become highly nonlinear,
with $f=e+ae_a$ being close to $\pm1$ for most values of $a$, and with
sharp transitions in between.  This means that the modal structures
consist of alternating regions in which either the periapses or the
apoapses are tightly bunched
(Fig.~\ref{f:mode_orbits_isothermal_rigid}).

\begin{figure*}
\includegraphics[width=\columnwidth]{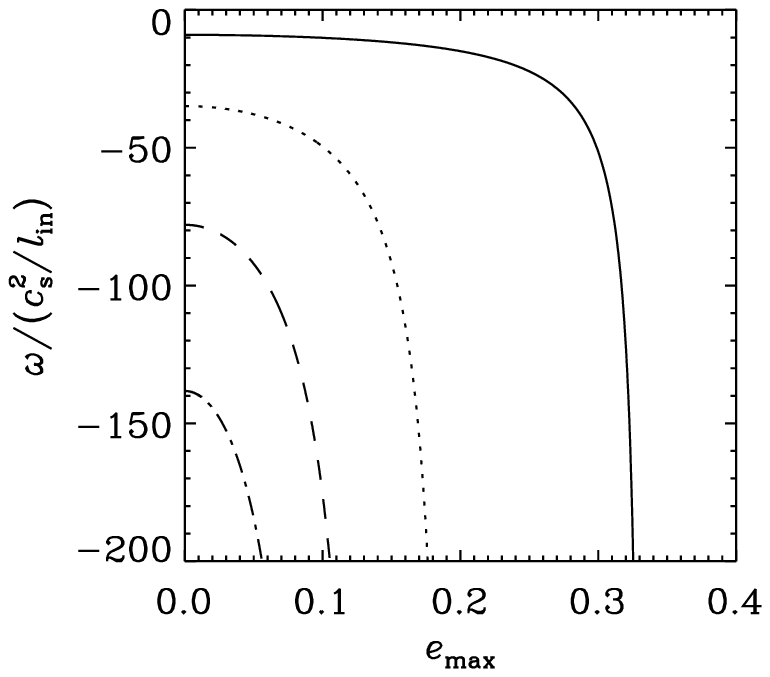}
\includegraphics[width=\columnwidth]{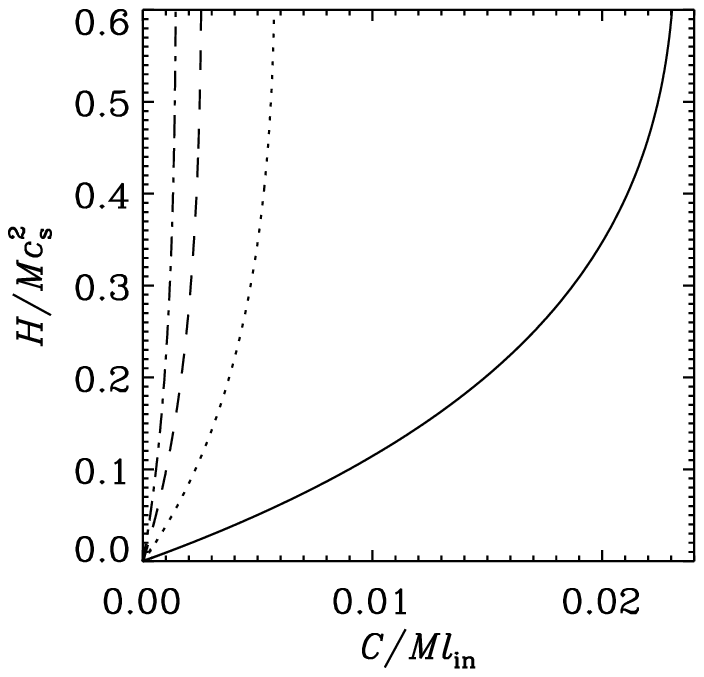}
\caption{\textit{Left}: dependence of the precession frequencies of
  the families of nonlinear eccentric modes (for the same problem as
  in Fig.~\ref{f:mode_shape_isothermal_rigid}) on the mode amplitude,
  quantified by $e_\mathrm{max}=\max|e|$.  Here
  $l_\mathrm{in}=\sqrt{GM_1r_\mathrm{in}}$ is the specific angular
  momentum of a circular orbit at the inner boundary.  \textit{Right}:
  Hamiltonian (energy) versus angular-momentum deficit for the same
  families.  The slope of each branch agrees with the negative of the
  precession frequency.  Note that each branch has a limiting value of
  $C$ at which $H\to\infty$ because an orbital intersection is
  approached.}
\label{f:mode_frequency_isothermal_rigid}
\end{figure*}

The variation of the precession rate $\omega$ with the mode amplitude
is shown in Fig.~\ref{f:mode_frequency_isothermal_rigid}, along with
the variation of the Hamiltonian with the angular-momentum
deficit. The precession is retrograde and increases with both
amplitude and mode number.  The relation $\rmd H/\rmd C=-\omega$
(i.e.\ $\rmd H/\rmd L=+\omega$) is verified numerically for each
branch.  The results for the lowest-order mode agree with those of
\citet{2016MNRAS.458.3739B}, which were obtained by solving a
nonlinear ODE for the eccentricity as a function of the semilatus
rectum.

When $e+ae_a=\pm1$ we have $ae=\pm a+\cst$.  The limiting form of the
lowest-order mode is given by
\begin{equation}
  ae=\begin{cases}
    a-r_\mathrm{in},&r_\mathrm{in}<a<\bar r,\\
    -a+r_\mathrm{out},&\bar r<a<r_\mathrm{out},
  \end{cases}
\end{equation}
where $\bar r=(r_\mathrm{in}+r_\mathrm{out})/2$.  It has a finite AMD
(equal to $0.02318\,Ml_\mathrm{in}$, where
$l_\mathrm{in}=\sqrt{GM_1r_\mathrm{in}}$, in the case we have
considered), which is the largest AMD that the system can support, and
an unbounded Hamiltonian because of the logarithmic divergence
(equation~\ref{f_gamma=1_untwisted}) as $f\to1$.  The limiting forms
of the higher-order modes, which were not considered by \citet{2016MNRAS.458.3739B}, have similar piecewise linear structures,
but the positions of the break points cannot be determined in a
straightforward way.

We next consider a more realistic problem involving an extended 2D
disc with $a_\mathrm{out}/a_\mathrm{in}=100$ and free boundaries.  We
also set $\gamma=2$ for this illustrative calculation, and in fact
consider a polytrope of index~$1$.  (As noted above, the geometric
Hamiltonian of a 2D disc can be obtained in terms of elementary
functions in the cases $\gamma=1$ and $\gamma=2$, which are expected
to bracket the range of behaviour expected in a real system.)  We take
$M_a/a\propto\varepsilon^\circ\propto T(a)/a$, where
$T(a)=\tanh[(a-a_\mathrm{in})/w_\mathrm{in}]\tanh[(a_\mathrm{out}-a)/w_\mathrm{out}]$
is a tapering function that is close to $1$ in most of the disc but
declines smoothly to $0$ at the inner and outer boundaries over
length-scales $w_\mathrm{in}=0.01a_\mathrm{in}$ and
$w_\mathrm{out}=0.01a_\mathrm{out}$.  Apart from the taper, this model
corresponds in the circular limit to a surface density
$\propto r^{-1}$ and a sound speed $\propto r^{-1/2}$.  The free
boundary conditions correspond to regularity conditions at
$a=a_\mathrm{in}$ and $a=a_\mathrm{out}$, which are singular points of
equation~(\ref{modal3}).  For this model they take the form
\begin{equation}
  2a_\mathrm{in}\f{\rmd T}{\rmd a}\f{\p F}{\p f}=\f{\omega na^2}{\varepsilon^\circ_\mathrm{in}}\f{e}{\sqrt{1-e^2}},
\end{equation}
where $\varepsilon^\circ_\mathrm{in}$ is the value of
$\varepsilon^\circ$ at $a=a_\mathrm{in}$ neglecting the taper; this
leads to a nonlinear relation between $e$ and $f$ at the boundaries.

\begin{figure*}
\includegraphics[width=\columnwidth]{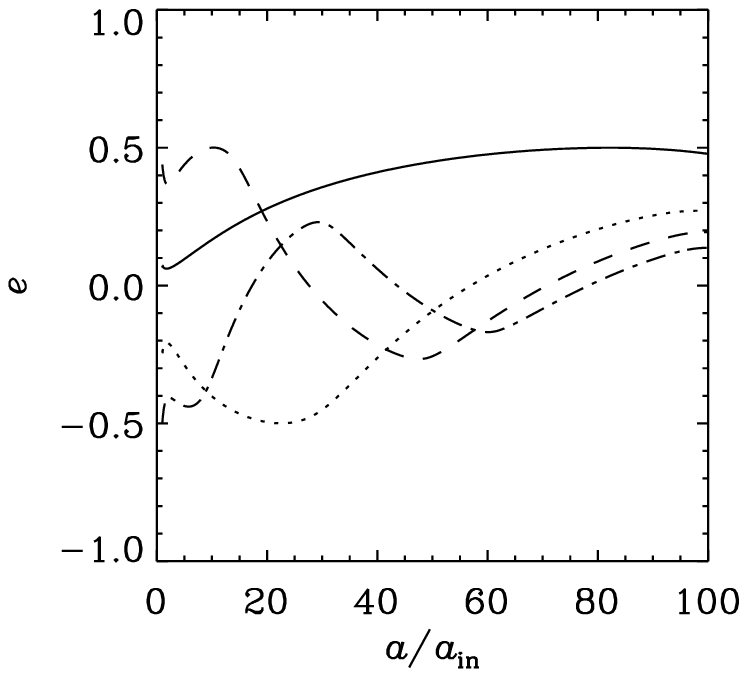}
\includegraphics[width=\columnwidth]{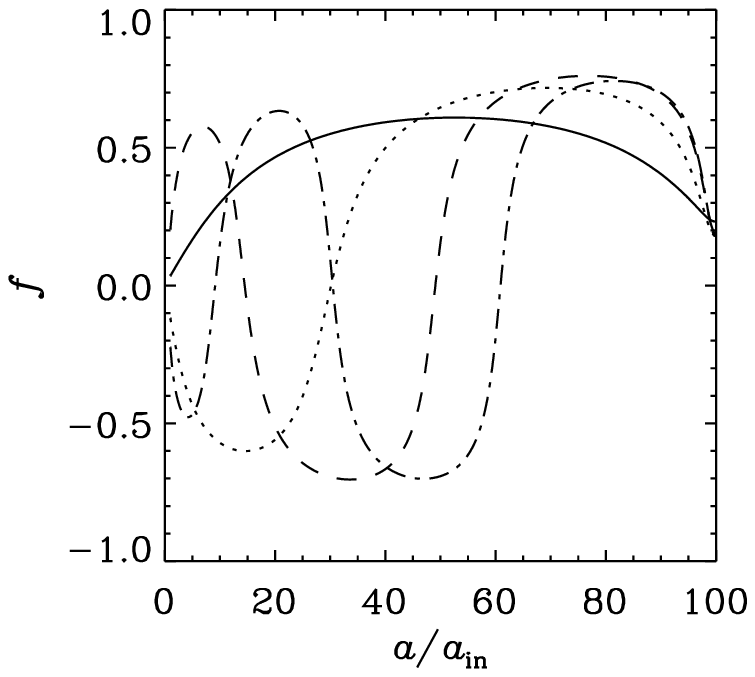}
\caption{Same as Fig.~\ref{f:mode_shape_isothermal_rigid} for the
  extended 2D disc with $\gamma=2$, $a_\mathrm{out}/a_\mathrm{in}=100$
  and free boundaries.  The amplitudes of the modes are chosen such
  that $e_\mathrm{max}=0.5$.}
\label{f:mode_shape_gamma=2_free_taper}
\end{figure*}

\begin{figure*}
\includegraphics[width=\columnwidth]{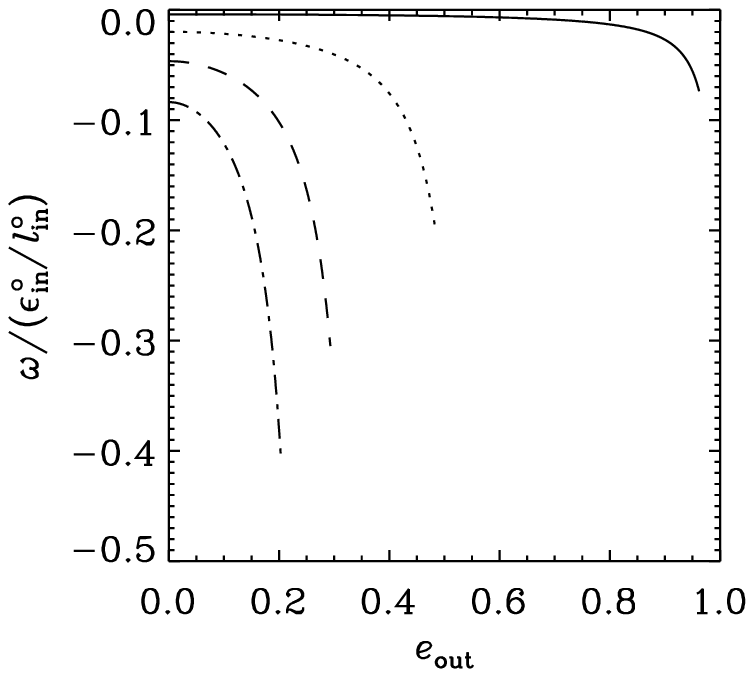}
\includegraphics[width=\columnwidth]{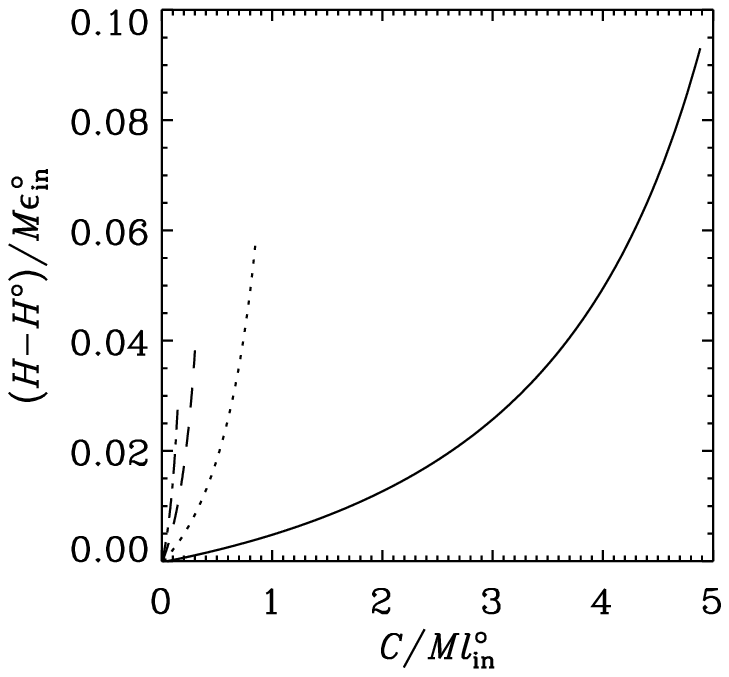}
\caption{Same as Fig.~\ref{f:mode_frequency_isothermal_rigid} for the
  extended 2D disc with $\gamma=2$, $a_\mathrm{out}/a_\mathrm{in}=100$
  and free boundaries.  Here $\varepsilon_\mathrm{in}^\circ$ is the specific
  internal energy of a circular disc at the inner boundary, neglecting
  the taper, and $l_\mathrm{in}^\circ=\sqrt{GM_1a_\mathrm{in}}$.  Apart from
  the lowest-order mode, the branches terminate when $e\to1$ at an
  interior point.}
\label{f:mode_frequency_gamma=2_free_taper}
\end{figure*}

The shapes of the first four modes are shown in
Fig.~\ref{f:mode_shape_gamma=2_free_taper}, where the amplitudes have
been chosen such a maximum eccentricity of $0.5$ is attained in each
case. Again, the modes form a sequence in which the number of nodes in
$e(a)$ increases by one. These modes are less confined than those of
the previous problem; with free boundary conditions, $e$ does not need
to vanish at the boundaries, and the broad extension of the disc means
that the eccentricity gradient can avoid the highly nonlinear limit
$f\to1$ even though $e$ is as large as $0.5$. The precession rate
again increases with amplitude and mode number and satisfies the
relation $\rmd H/\rmd C=-\omega$
(Fig.~\ref{f:mode_frequency_gamma=2_free_taper}). The numerical values
of $-\omega$ in units of $\varepsilon_\mathrm{in}/l_\mathrm{in}$ are
much smaller than the values of $-\omega$ in units of
$c_\rms/l_\mathrm{in}$ found in the previous problem
(Fig.~\ref{f:mode_frequency_isothermal_rigid}). This is because (i)
the eccentricity gradients are more moderate, which reduces the
precession rate; (ii) the internal energy in most of the disc is
significantly less than $\varepsilon_\mathrm{in}$ because of the
assumed temperature gradient; and (iii) the majority of the angular
momentum is located at $a\gg a_\mathrm{in}$.

\section{Conclusion}
\label{s:conclusion}

In this paper we have developed a new framework for studying the
nonlinear dynamics of eccentric astrophysical discs.  In the case of
an ideal, non-self-gravitating fluid, a thin eccentric disc consists
of a continuous set of nested elliptic orbits that evolve slowly as a
result of pressure.  The semimajor axis of each orbit is conserved, as
is the total angular-momentum deficit of the disc, while the orbits
undergo precession and exchanges of eccentricity.  The nonlinear
dynamics of this system can be derived from a convenient variational
principle and expressed in a concise Hamiltonian form, using either
canonical (e.g.\ modified Delaunay) variables or the more familiar
Keplerian orbital elements.  The Hamiltonian of a 2D disc is just the
orbit-averaged internal energy of the disc, which can be determined
from its eccentricity distribution using the geometry of the
elliptical orbits.  In the realistic case of a 3D disc, the
Hamiltonian needs to be modified to take into account the dynamical
vertical structure of the disc, but is still equal to a multiple of a
suitably averaged internal energy.

The simplest solutions of the theory are uniformly precessing
nonlinear eccentric modes, which make the Hamiltonian stationary
subject to the angular momentum deficit being fixed.  We have
presented numerical examples of branches of modal solutions in
illustrative situations up to their maximum possible amplitudes.

In many astronomical situations of interest, eccentric discs are
expected to interact gravitationally with orbital companions; examples
include circumbinary discs around young binary stars or binary black
holes in galactic nuclei, circumstellar discs in cataclysmic variables
and young binary stars, and protoplanetary discs with gap-opening
planets. These interactions generally involve both secular effects,
which contribute to precession and exchanges of eccentricity between
the disc and the companion, and resonant effects, which contribute to
the growth or damping of eccentricity. Both have been considered in a
linear regime by \citet{2006MNRAS.368.1123G} and
\citet{2016MNRAS.458.3221T}. It is likely that the Hamiltonian
formalism developed here will be useful in understanding the nonlinear
dynamics found in long-term simulations by \citet{2017MNRAS.464L.114R}
and \citet{2018MNRAS.474.4460R}, as it is a description of an ideal
fluid that connects more readily with the established methods of
celestial mechanics.  It may also be helpful in understanding
instabilities of eccentric discs that involve mode couplings
\citep{2008MNRAS.386.2297F,2008PASJ...60..111K,2014MNRAS.445.2637B}.
It would be valuable in future to explore further the nonlinear
dynamics of 3D eccentric discs using this approach.

The theory developed here is intended to be easier to understand and
to use than the nonlinear description of eccentric discs worked out by
\citet{2001MNRAS.325..231O}.  The older work considers the
eccentricity as a function of semilatus rectum (or angular momentum)
rather than semimajor axis (or energy), and requires a number of
stress integrals to be computed.  The new theory is all derived from a
scalar Hamiltonian but does not include dissipative effects.  It would
be valuable to try to incorporate the effects of viscous dissipation,
heating and cooling while preserving as much as possible of the
unifying structure of this theory.

\section*{Acknowledgements}

This research was supported by STFC through grants ST/L000636/1 and
ST/P000673/1.

\appendix

\onecolumn

\section{Derivation of the Hamiltonian form of the equations}
\label{s:appendixa}

In the Hamiltonian theory of dynamics, any canonical transformation
from phase-space variables $(\bmq,\bmp)$ to $(\bmQ,\bmP)$ satisfies
the `direct conditions'
\begin{equation}
  \f{\p P_i}{\p p_j}=\f{\p q_j}{\p Q_i},\qquad
  \f{\p P_i}{\p q_j}=-\f{\p p_j}{\p Q_i},\qquad
  \f{\p Q_i}{\p p_j}=-\f{\p q_j}{\p P_i},\qquad
  \f{\p Q_i}{\p q_j}=\f{\p p_j}{\p P_i}.
\end{equation}
We apply these relations with $(\bmq,\bmp)$ being the Cartesian
phase-space coordinates $(x,y,v_x,v_y)$ and $(\bmQ,\bmP)$ being the
action--angle variables $(\lambda,\gamma,\Lambda,\Gamma)$.  We use the
special symbol $\eth$ (rather than $\partial$) to denote partial
derivatives of the four-dimensional phase-space transformation
$(x,y,v_x,v_y)\mapsto(\lambda,\gamma,\Lambda,\Gamma)$ in order to
distinguish them from those of the two-dimensional real-space
transformation $(x,y)\mapsto(\Lambda,\lambda)$ associated with the
orbital coordinate system.  Thus, in particular,
\begin{equation}
  \f{\eth\Lambda}{\eth v_x}=\f{\eth x}{\eth\lambda},\qquad
  \f{\eth\Lambda}{\eth v_y}=\f{\eth y}{\eth\lambda},\qquad
  \f{\eth\Gamma}{\eth v_x}=\f{\eth x}{\eth\gamma},\qquad
  \f{\eth\Gamma}{\eth v_y}=\f{\eth y}{\eth\gamma},\qquad
  \f{\eth\gamma}{\eth v_x}=-\f{\eth x}{\eth\Gamma},\qquad
  \f{\eth\gamma}{\eth v_y}=-\f{\eth y}{\eth\Gamma}.
\end{equation}

For a planar system, in the presence of a perturbing force per unit
mass $\bmf$, the equation of motion of a fluid element or test
particle is
\begin{equation}
  \dot v_x=\dots+f_x,\qquad
  \dot v_y=\dots+f_y,
\end{equation}
where the dots represent the Newtonian force due to the central
mass. Using the chain rule and the direct conditions, we deduce that
the canonical variables that would be constant in a Keplerian orbit
evolve according to
\begin{align}
&  \dot\Lambda=\f{\eth\Lambda}{\eth v_x}f_x+\f{\eth\Lambda}{\eth v_y}f_y=\f{\eth x}{\eth\lambda}f_x+\f{\eth y}{\eth\lambda}f_y,\\
&  \dot\Gamma=\f{\eth\Gamma}{\eth v_x}f_x+\f{\eth\Gamma}{\eth v_y}f_y=\f{\eth x}{\eth\gamma}f_x+\f{\eth y}{\eth\gamma}f_y,\\
&  \dot\gamma=\f{\eth\gamma}{\eth v_x}f_x+\f{\eth\gamma}{\eth v_y}f_y=-\f{\eth x}{\eth\Gamma}f_x-\f{\eth y}{\eth\Gamma}f_y.
\end{align}
(We will not require $\dot\lambda$, which in any case is dominated by
the Keplerian mean motion $n$.)

The partial derivatives of the 2D transformation are related to those
of the 4D transformation by
\begin{equation}
  \f{\p x}{\p\Lambda}=\f{\eth x}{\eth\Lambda}+\f{\eth x}{\eth\Gamma}\Gamma_\Lambda+\f{\eth x}{\eth\gamma}\gamma_\Lambda,\qquad
  \f{\p x}{\p\lambda}=\f{\eth x}{\eth\lambda},\qquad
  \f{\p y}{\p\Lambda}=\f{\eth y}{\eth\Lambda}+\f{\eth y}{\eth\Gamma}\Gamma_\Lambda+\f{\eth y}{\eth\gamma}\gamma_\Lambda,\qquad
  \f{\p y}{\p\lambda}=\f{\eth y}{\eth\lambda}.
\end{equation}
The Jacobian matrix of the 2D transformation is
\begin{equation}
  \begin{pmatrix}
    \p x/\p\Lambda&\p x/\p\lambda\\
    \p y/\p\Lambda&\p y/\p\lambda
  \end{pmatrix},
\end{equation}
with determinant
\begin{equation}
  J=\f{\p x}{\p\Lambda}\f{\p y}{\p\lambda}-\f{\p x}{\p\lambda}\f{\p y}{\p\Lambda}=\left(\f{\eth x}{\eth\Lambda}\f{\eth y}{\eth\lambda}-\f{\eth x}{\eth\lambda}\f{\eth y}{\eth\Lambda}\right)+\left(\f{\eth x}{\eth\Gamma}\f{\eth y}{\eth\lambda}-\f{\eth x}{\eth\lambda}\f{\eth y}{\eth\Gamma}\right)\Gamma_\Lambda+\left(\f{\eth x}{\eth\gamma}\f{\eth y}{\eth\lambda}-\f{\eth x}{\eth\lambda}\f{\eth y}{\eth\gamma}\right)\gamma_\Lambda
\label{j}
\end{equation}
and inverse
\begin{equation}
  \begin{pmatrix}
    \p\Lambda/\p x&\p\Lambda/\p y\\
    \p\lambda/\p x&\p\lambda/\p y
  \end{pmatrix}=
  \f{1}{J}
  \begin{pmatrix}
    \p y/\p\lambda&-\p x/\p\lambda\\
    -\p y/\p\Lambda&\p x/\p\Lambda
  \end{pmatrix}.
\end{equation}

For a perturbing force of the form
\begin{equation}
  \bmf=-\f{1}{\Sigma}\grad P,
\end{equation}
where $P$ is a vertically integrated pressure (or other isotropic
stress), we deduce that
\begin{equation}
  \Sigma\dot\Lambda=-\f{\eth x}{\eth\lambda}\f{\p P}{\p x}-\f{\eth y}{\eth\lambda}\f{\p P}{\p y},\qquad
  \Sigma\dot\Gamma=-\f{\eth x}{\eth\gamma}\f{\p P}{\p x}-\f{\eth y}{\eth\gamma}\f{\p P}{\p y},\qquad
  \Sigma\dot\gamma=\f{\eth x}{\eth\Gamma}\f{\p P}{\p x}+\f{\eth y}{\eth\Gamma}\f{\p P}{\p y}.
\end{equation}
Using the chain rule on the partial derivatives of $P$ gives
\begin{align}
&  \Sigma\dot\Lambda=-\f{\p P}{\p\lambda},\\
&  \Sigma\dot\Gamma=-\left(\f{\eth x}{\eth\gamma}\f{\p\Lambda}{\p x}+\f{\eth y}{\eth\gamma}\f{\p\Lambda}{\p y}\right)\f{\p P}{\p\Lambda}-\left(\f{\eth x}{\eth\gamma}\f{\p\lambda}{\p x}+\f{\eth y}{\eth\gamma}\f{\p\lambda}{\p y}\right)\f{\p P}{\p\lambda},\\
&  \Sigma\dot\gamma=\left(\f{\eth x}{\eth\Gamma}\f{\p\Lambda}{\p x}+\f{\eth y}{\eth\Gamma}\f{\p\Lambda}{\p y}\right)\f{\p P}{\p\Lambda}+\left(\f{\eth x}{\eth\Gamma}\f{\p\lambda}{\p x}+\f{\eth y}{\eth\Gamma}\f{\p\lambda}{\p y}\right)\f{\p P}{\p\lambda}.
\end{align}
Multiplying by $J$ and using the inverse Jacobian matrix gives
\begin{align}
&  J\Sigma\dot\Lambda=-J\f{\p P}{\p\lambda},\\
&  J\Sigma\dot\Gamma=-\left(\f{\eth x}{\eth\gamma}\f{\p y}{\p\lambda}-\f{\eth y}{\eth\gamma}\f{\p x}{\p\lambda}\right)\f{\p P}{\p\Lambda}-\left(-\f{\eth x}{\eth\gamma}\f{\p y}{\p\Lambda}+\f{\eth y}{\eth\gamma}\f{\p x}{\p\Lambda}\right)\f{\p P}{\p\lambda},\\
&  J\Sigma\dot\gamma=\left(\f{\eth x}{\eth\Gamma}\f{\p y}{\p\lambda}-\f{\eth y}{\eth\Gamma}\f{\p x}{\p\lambda}\right)\f{\p P}{\p\Lambda}+\left(-\f{\eth x}{\eth\Gamma}\f{\p y}{\p\Lambda}+\f{\eth y}{\eth\Gamma}\f{\p x}{\p\Lambda}\right)\f{\p P}{\p\lambda}.
\end{align}
Now we notice from equation~(\ref{j}) that
\begin{equation}
  \f{\p J}{\p\Gamma_\Lambda}=\f{\eth x}{\eth\Gamma}\f{\eth y}{\eth\lambda}-\f{\eth x}{\eth\lambda}\f{\eth y}{\eth\Gamma},\qquad
  \f{\p J}{\p\gamma_\Lambda}=\f{\eth x}{\eth\gamma}\f{\eth y}{\eth\lambda}-\f{\eth x}{\eth\lambda}\f{\eth y}{\eth\gamma}.
\end{equation}
Furthermore, we have the differential identities
\begin{equation}
  \f{\p}{\p\lambda}\left(\f{\eth x}{\eth\Gamma}\f{\p y}{\p\Lambda}-\f{\eth y}{\eth\Gamma}\f{\p x}{\p\Lambda}\right)=\f{\p}{\p\Lambda}\left(\f{\p J}{\p\Gamma_\Lambda}\right)-\f{\p J}{\p\Gamma},\qquad
  \f{\p}{\p\lambda}\left(\f{\eth x}{\eth\gamma}\f{\p y}{\p\Lambda}-\f{\eth y}{\eth\gamma}\f{\p x}{\p\Lambda}\right)=\f{\p}{\p\Lambda}\left(\f{\p J}{\p\gamma_\Lambda}\right)-\f{\p J}{\p\gamma},
\end{equation}
which do not rely on any special properties of the functions
$x(\lambda,\gamma,\Lambda,\Gamma)$ and
$y(\lambda,\gamma,\Lambda,\Gamma)$.
We now carry out an orbital time-average of the equations, denoted by
angle brackets:
\begin{equation}
  \langle\cdot\rangle=\f{1}{2\pi}\int_0^{2\pi}\cdot\,\rmd\lambda.
\end{equation}
This averaging operation, which also corresponds to a mass-weighted
spatial average around the orbit, allows us to integrate by parts with
respect to the periodic variable $\lambda$, with the results
\begin{align}
&  J\Sigma\langle\dot\Lambda\rangle=\left\langle P\f{\p J}{\p\lambda}\right\rangle,\\
&  J\Sigma\langle\dot\Gamma\rangle=-\f{\p}{\p\Lambda}\left\langle P\f{\p J}{\p\gamma_\Lambda}\right\rangle+\left\langle P\f{\p J}{\p\gamma}\right\rangle,\\
&  J\Sigma\langle\dot\gamma\rangle=\f{\p}{\p\Lambda}\left\langle P\f{\p J}{\p\Gamma_\Lambda}\right\rangle-\left\langle P\f{\p J}{\p\Gamma}\right\rangle.
\end{align}

\subsection{2D analysis}

In the artificial but widely considered case of a 2D ideal fluid,
variations of the vertically integrated pressure $P$ and the specific
internal energy $\varepsilon$ around each orbit are adiabatic and
related by $\rmd\varepsilon=-P\,\rmd A$, where
$A=1/\Sigma=(2\pi/M_\Lambda)J$ is the specific area.  The evolutionary
equations, multiplied by $2\pi$, then reduce to
\begin{align}
&  M_\Lambda\langle\dot\Lambda\rangle=-M_\Lambda\left\langle\f{\p\varepsilon}{\p\lambda}\right\rangle=0,\\
&  M_\Lambda\langle\dot\Gamma\rangle=\f{\p}{\p\Lambda}\left(M_\Lambda\left\langle\f{\p\varepsilon}{\p\gamma_\Lambda}\right\rangle\right)-M_\Lambda\left\langle\f{\p\varepsilon}{\p\gamma}\right\rangle,\\
&  M_\Lambda\langle\dot\gamma\rangle=-\f{\p}{\p\Lambda}\left(M_\Lambda\left\langle\f{\p\varepsilon}{\p\Gamma_\Lambda}\right\rangle\right)+M_\Lambda\left\langle\f{\p\varepsilon}{\p\Gamma}\right\rangle.
\end{align}
The first equation implies that $\Lambda$, and therefore $a$, is a
material invariant in this theory, which is true because there is no
dissipation of energy, nor does the pressure transfer any energy
between neighbouring orbits.  The three equations are in the canonical
Hamiltonian form
\begin{equation}
  M_\Lambda\langle\dot\Lambda\rangle=-\f{\delta H}{\delta\lambda}=0,\qquad
  M_\Lambda\langle\dot\Gamma\rangle=-\f{\delta H}{\delta\gamma},\qquad
  M_\Lambda\langle\dot\gamma\rangle=\f{\delta H}{\delta\Gamma},
\end{equation}
where the Hamiltonian $H$ is given by
\begin{equation}
  H=\int H_\Lambda\,\rmd\Lambda
\end{equation}
with Hamiltonian density
\begin{equation}
  H_\Lambda=M_\Lambda\langle\varepsilon\rangle
\end{equation}
involving the orbit-averaged specific internal energy
$\langle\varepsilon\rangle$.

\subsection{3D analysis}

When the third ($z$) dimension perpendicular to the plane of the disc
is considered, it can be seen that eccentric discs are never in
hydrostatic equilibrium in this vertical direction.  A non-zero
eccentricity leads to an oscillatory variation of the vertical
gravitational acceleration around the orbit, while a non-zero
eccentricity gradient or twist leads to an oscillatory horizontal
compression of the fluid.  In either case the disc undergoes a forced
vertical `breathing mode'
\citep{2001MNRAS.325..231O,2014MNRAS.445.2621O}.  The differential
equation describing the dependence of the vertical
scaleheight\footnote{There is an unfortunate notational clash between
  the scaleheight and the Hamiltonian.  The context should make it
  clear which meaning $H$ has in any equation.} $H$ around the orbit,
for an ideal fluid disc that is stationary in an inertial frame on the
orbital timescale, can be written as
\begin{equation}
  n^2\f{\p^2H}{\p\lambda^2}+\Psi H=\f{P}{\Sigma H},
\label{odeH}
\end{equation}
where
\begin{equation}
  \Psi=\f{\p^2\Phi}{\p z^2}\bigg|_{z=0}=\f{GM}{r^3}
\end{equation}
describes the vertical gravity in a disc (assumed here to be thin),
while $P$ and $\Sigma$ are the vertically integrated pressure and
density.  (Equation~\ref{odeH} is equivalent to equations 141--142 of
\citet{2014MNRAS.445.2621O}.)  The mass-weighted vertically averaged
specific internal energy of a perfect gas is
\begin{equation}
  \bar\varepsilon=\int\f{p}{(\gamma-1)\rho}\,\rho\,\rmd z\bigg/\int\rho\,\rmd z=\f{P}{(\gamma-1)\Sigma}.
\end{equation}
Adiabatic flow means that $(P/H)\propto(\Sigma/H)^\gamma$ around the
orbit, so $\rmd\bar\varepsilon=-(P/H)\,\rmd(AH)$ when $\Lambda$ is
held constant.

The associated energy equation is obtained by multiplying
equation~(\ref{odeH}) by $\p H/\p\lambda$:
\begin{equation}
  \f{\p}{\p\lambda}\left[\f{1}{2}n^2\left(\f{\p H}{\p\lambda}\right)^2+\f{1}{2}\Psi H^2+\bar\varepsilon\right]=\f{1}{2}\f{\p\Psi}{\p\lambda}H^2-P\f{\p A}{\p\lambda}.
\end{equation}
The three terms in the square brackets are the kinetic, gravitational
and internal energies per unit mass, after vertical averaging.  The two
source terms on the right-hand side are due to variable vertical
gravity and horizontal orbital compression, respectively.  An orbital
average gives
\begin{equation}
  \left\langle\f{1}{2}\f{\p\Psi}{\p\lambda}H^2-P\f{\p A}{\p\lambda}\right\rangle=0.
\label{ave1}
\end{equation}
On the other hand, multiplying equation~(\ref{odeH}) by $H$ and
averaging (with integration by parts) gives the virial relation
\begin{equation}
  \left\langle-n^2\left(\f{\p H}{\p\lambda}\right)^2+\Psi H^2-(\gamma-1)\bar\varepsilon\right\rangle=0.
\label{virial}
\end{equation}

Now consider how the scaleheight responds to variations in the
geometry of the disc.  Note that, of the two agents driving the
vertical oscillation, $\Psi$ depends on $(\Gamma,\gamma)$ as well as
$(\Lambda,\lambda)$, while $A$ also depends on $\Gamma_\Lambda$ and
$\gamma_\Lambda$.  Consider, for example, variations of $\Gamma$.  We
differentiate equation~(\ref{odeH}) with respect to $\Gamma$ to obtain
\begin{equation}
  n^2\f{\p^3H}{\p\lambda^2\p\Gamma}+\f{\p\Psi}{\p\Gamma}H+\Psi\f{\p H}{\p\Gamma}=\f{P}{\Sigma H}\left[-(\gamma-1)\f{\p\ln A}{\p\Gamma}-\gamma\f{\p\ln H}{\p\Gamma}\right].
\end{equation}
We then multiply by $H$ and average (with integration by parts):
\begin{equation}
  \left\langle-n^2\f{\p^2H}{\p\lambda\p\Gamma}\f{\p H}{\p\lambda}+\f{\p\Psi}{\p\Gamma}H^2+\Psi\f{\p H}{\p\Gamma}H\right\rangle=-(\gamma-1)\left\langle P\f{\p A}{\p\Gamma}\right\rangle-\left\langle\f{\gamma P}{\Sigma H}\f{\p H}{\p\Gamma}\right\rangle.
\end{equation}
Using the property
\begin{equation}
  \f{\p\bar\varepsilon}{\p\Gamma}=-\f{P}{H}\f{\p(AH)}{\p\Gamma},
\end{equation}
we can rearrange to obtain
\begin{equation}
  \left\langle\f{1}{2}\f{\p\Psi}{\p\Gamma}H^2-P\f{\p A}{\p\Gamma}\right\rangle=\f{\p\mathcal{H}}{\p\Gamma},
\label{ave2}
\end{equation}
where
\begin{equation}
  \mathcal{H}=\left\langle\f{1}{2}n^2\left(\f{\p H}{\p\lambda}\right)^2-\f{1}{2}\Psi H^2+\gamma\bar\varepsilon\right\rangle.
\end{equation}
Using the virial relation, this quantity can be rewritten either as
\begin{equation}
  \mathcal{H}=\f{1}{2}(\gamma+1)\langle\bar\varepsilon\rangle,
\end{equation}
which is a multiple of the average internal energy, or as 
\begin{equation}
  \mathcal{H}=\left\langle-\f{1}{2}n^2\left(\f{\p H}{\p\lambda}\right)^2+\f{1}{2}\Psi H^2+\bar\varepsilon\right\rangle,
\end{equation}
which is the average of (potential--kinetic) energy.  Indeed, this is
(minus) the averaged Lagrangian that appears in the theory of
nonlinear dispersive waves \citep{1965JFM....22..273W}.

It can be shown similarly that
\begin{equation}
  \left\langle\f{1}{2}\f{\p\Psi}{\p\gamma}H^2-P\f{\p A}{\p\gamma}\right\rangle=\f{\p\mathcal{H}}{\p\gamma}.
\label{ave3}
\end{equation}

When the additional horizontal gravitational force (with mass-weighted
vertical averaging) due to $\Psi$ is added to the evolutionary
equations, we have
\begin{align}
&  J\Sigma\langle\dot\Lambda\rangle=\left\langle P\f{\p J}{\p\lambda}\right\rangle-J\Sigma\left\langle\f{1}{2}\f{\p\Psi}{\p\lambda}H^2\right\rangle,\\
&  J\Sigma\langle\dot\Gamma\rangle=-\f{\p}{\p\Lambda}\left\langle P\f{\p J}{\p\gamma_\Lambda}\right\rangle+\left\langle P\f{\p J}{\p\gamma}\right\rangle-J\Sigma\left\langle\f{1}{2}\f{\p\Psi}{\p\gamma}H^2\right\rangle,\\
&  J\Sigma\langle\dot\gamma\rangle=\f{\p}{\p\Lambda}\left\langle P\f{\p J}{\p\Gamma_\Lambda}\right\rangle-\left\langle P\f{\p J}{\p\Gamma}\right\rangle+J\Sigma\left\langle\f{1}{2}\f{\p\Psi}{\p\Gamma}H^2\right\rangle.
\end{align}
Using equations (\ref{ave1}), (\ref{ave2}) and (\ref{ave3}), we see
that $\langle\dot\Lambda\rangle$ evaluates to zero, and that the
equations are of Hamiltonian form, with Hamiltonian density given by
\begin{equation}
  H_\Lambda=M_\Lambda\mathcal{H}=\f{1}{2}(\gamma+1)M_\Lambda\langle\bar\varepsilon\rangle.
\end{equation}

\section{Dimensionless forms of the Hamiltonian density}
\label{s:appendixb}

\subsection{2D analysis}

For a perfect gas of adiabatic index $\gamma$,
adiabatic flow around each orbit means that $P=K\Sigma^\gamma$, where
$K(a)$ is related to the entropy distribution, which does not evolve.
The specific internal energy is then
\begin{equation}
  \varepsilon=\f{P}{(\gamma-1)\Sigma}=\f{K\Sigma^{\gamma-1}}{\gamma-1}.
\end{equation}
Recall that equation~(\ref{sigma}) relates the surface density
$\Sigma$ to the mass distribution $M_\Lambda$ and the Jacobian
$J=J^\circ j$.  If the disc had the same distributions of mass and
entropy but were circular, it would have
$\Sigma=\Sigma^\circ=M_\Lambda/2\pi J^\circ$ and
$P=P^\circ=K(\Sigma^\circ)^\gamma$.  Thus $\Sigma=\Sigma^\circ j^{-1}$ and
\begin{equation}
  H_a=M_a\langle\varepsilon\rangle=H^\circ_a F^\mathrm{(2D)},
\end{equation}
where
\begin{equation}
  H^\circ_a=2\pi aP^\circ
\end{equation}
is a given function of $a$ and
\begin{align}
  F^\mathrm{(2D)}&=\f{1}{\gamma-1}\left\langle j^{-(\gamma-1)}\right\rangle\nonumber\\
  &=\left(\f{1}{\gamma-1}\right)\f{1}{2\pi}\int_0^{2\pi}\left[\f{1-e(e+ae_a)}{\sqrt{1-e^2}}-\f{ae_a\cos E}{\sqrt{1-e^2}}-ae\varpi_a\sin E\right]^{-(\gamma-1)}(1-e\cos E)\,\rmd E
\label{f2d}
\end{align}
is the dimensionless, geometric part of the Hamiltonian density, which
depends on $e$, $ae_a$, $ae\varpi_a$ and $\gamma$.

In the case of a (globally) isothermal disc we instead have
\begin{equation}
  \varepsilon=c_\rms^2\ln\Sigma+\cst,
\end{equation}
where $c_\rms=\cst$ is the isothermal sound speed.  Then
\begin{equation}
  H_a=M_a\langle\varepsilon\rangle=H^\circ_a\left(\ln\Sigma^\circ+\cst+F^\mathrm{(2D)}\right),
\label{h_lambda_iso}
\end{equation}
where
\begin{equation}
  H^\circ_a=M_ac_\rms^2=2\pi aP^\circ
\end{equation}
and
\begin{equation}
  F^\mathrm{(2D)}=-\langle\ln j\rangle=-\f{1}{2\pi}\int_0^{2\pi}\ln\left[\f{1-e(e+ae_a)}{\sqrt{1-e^2}}-\f{ae_a\cos E}{\sqrt{1-e^2}}-ae\varpi_a\sin E\right](1-e\cos E)\,\rmd E.
\label{f2d_iso}
\end{equation}
This expression agrees with the limit $\gamma\to1$ of
equation~(\ref{f2d}) after the constant term $1/(\gamma-1)$ (which, like
the terms depending only on $a$ in equation~\ref{h_lambda_iso}, does
not affect the dynamics) is removed from it.  This follows from the
result
\begin{equation}
  \lim_{\gamma\to1}\f{j^{-(\gamma-1)}-1}{\gamma-1}=-\ln j.
\end{equation}

\subsection{3D analysis}

For a 3D disc, adiabatic flow means instead that
$(P/H)=K(\Sigma/H)^\gamma$, where $K$ is a fixed function of $a$.  If
we write the scaleheight as $H=H^\circ(a)h(E)$, where
$H^\circ=(P^\circ/\Sigma^\circ n^2)^{1/2}$ is the (hydrostatic)
vertical scaleheight of a circular disc with the same mass and entropy
distributions, then we have $\Sigma=\Sigma^\circ j^{-1}$ and
$P=P^\circ j^{-\gamma}h^{-(\gamma-1)}$.  Equation~(\ref{odeH}) reduces
to a dimensionless ODE for the dimensionless scaleheight $h$:
\begin{equation}
  \f{\rmd^2h}{\rmd\lambda^2}+\f{h}{(1-e\cos E)^3}=\f{1}{j^{\gamma-1}h^\gamma},
\end{equation}
or, in terms of derivatives with respect to the eccentric anomaly,
\begin{equation}
  (1-e\cos E)\f{\rmd^2h}{\rmd E^2}-e\sin E\f{\rmd h}{\rmd E}+h=\f{(1-e\cos E)^3}{j^{\gamma-1}h^\gamma}.
\label{ode}
\end{equation}
We then have the Hamiltonian density
\begin{equation}
  H_a=\f{1}{2}(\gamma+1)M_a\langle\bar\varepsilon\rangle=H_a^\circ F^\mathrm{(3D)},
\end{equation}
with
\begin{equation}
  H_a^\circ=2\pi aP^\circ
\end{equation}
and
\begin{equation}
  F^\mathrm{(3D)}=\f{(\gamma+1)}{2(\gamma-1)}\left\langle(jh)^{-(\gamma-1)}\right\rangle.
\end{equation}

In the isothermal case we have instead
\begin{equation}
  \bar\varepsilon=c_\rms^2\ln\left(\f{\Sigma}{H}\right)+\cst,
\end{equation}
\begin{equation}
  H_a=M_a\langle\bar\varepsilon\rangle=H_a^\circ\left(\ln\Sigma^\circ-\ln H^\circ+\cst+F^\mathrm{(3D)}\right),
\end{equation}
\begin{equation}
  F^\mathrm{(3D)}=-\langle\ln(jh)\rangle.
\end{equation}

\section{Analaytical expressions for the Hamiltonian of a 2D disc}
\label{s:appendixc}

The quantity $j$ in square brackets in equation (\ref{f2d}) or
(\ref{f2d_iso}) can be written as
\begin{equation}
  j=\f{1-e(e+ae_a)}{\sqrt{1-e^2}}(1-q\cos\theta),
\label{jtheta}
\end{equation}
where $\theta=E-\alpha$, with the amplitude $q$ and phase $\alpha$
satisfying
\begin{equation}
  q\cos\alpha=\f{ae_a}{1-e(e+ae_a)},\qquad
  q\sin\alpha=\f{\sqrt{1-e^2}\,ae\varpi_a}{1-e(e+ae_a)},
\end{equation}
which implies
\begin{equation}
  q^2=\f{(ae_a)^2+(1-e^2)(ae\varpi_a)^2}{[1-e(e+ae_a)]^2},\qquad
  1-q^2=\f{(1-e^2)[1-(e+ae_a)^2-(ae\varpi_a)^2]}{[1-e(e+ae_a)]^2},\qquad
  \tan\alpha=\f{\sqrt{1-e^2}\,e\varpi_a}{e_a}.
\end{equation}
To avoid orbital intersection we require $|q|<1$.  In the untwisted
case $\varpi_a=0$ it is natural to take $\alpha=0$ and regard $q$ as a
signed quantity.  $q$ is a measure of the degree of areal compression
and the level of nonlinearity in an eccentric disc; it generalizes the
nonlinearity parameter used in the theory of planetary rings in the
regime $e\ll1$.

For general $\gamma$ the integral~(\ref{f2d}) can be evaluated in
terms of Legendre functions.  The special case $\gamma=2$ and the
limiting isothermal case $\gamma=1$ involve only elementary functions.

Let
\begin{equation}
  I_p(q)=\f{1}{2\pi}\int_0^{2\pi}\f{\rmd\theta}{(1-q\cos\theta)^p}
\end{equation}
for $|q|<1$. This is an even function of $q$ with $I_p(0)=1$. For
$p>0$ or $p<-1$ it is a monotonically increasing function of $q$ for
$0<q<1$. For $p>1/2$ it diverges proportionally to $(1-q)^{(1/2)-p}$
as $q\to1$. For $-1<p<0$ it is monotonically decreasing. For $p=0$ or
$p=-1$ it is constant and equal to $1$. Special cases include
\begin{equation}
  I_{-1}(q)=1,\qquad
  I_0(q)=1,\qquad
  I_1(q)=(1-q^2)^{-1/2},\qquad
  I_2(q)=(1-q^2)^{-3/2}.
\end{equation}
In general, $I_p(q)$ can be written in terms of the Legendre function:
\begin{equation}
  I_p(q)=z^pP_{p-1}(z),\qquad
  z=\f{1}{\sqrt{1-q^2}}.
\end{equation}
A series expansion, convergent for all $|q|<1$, is
\begin{equation}
  I_p(q)=\sum_{n=0}^\infty\binom{p+2n-1}{2n}J_nq^{2n},
\end{equation}
where
\begin{equation}
  J_n=\f{1}{2\pi}\int_0^{2\pi}\cos^{2n}\theta\,\rmd\theta=\f{(2n)!}{2^{2n}(n!)^2},
\end{equation}
which behaves as
\begin{equation}
  J_n\sim\f{1}{\sqrt{\pi n}}\qquad\hbox{as $n\to\infty$}.
\end{equation}
Thus
\begin{equation}
  I_p(q)=\sum_{n=0}^\infty\f{(p+2n-1)(p+2n-2)\cdots p}{2^{2n}(n!)^2}\,q^{2n}.
\end{equation}

To express the integral~(\ref{f2d}) in terms of $I_p(q)$, we use
equation~(\ref{jtheta}) and change variables from $E$ to $\theta$.
Note that
\begin{equation}
  \f{1}{2\pi}\int_0^{2\pi}\f{\cos\theta\,\rmd\theta}{(1-q\cos\theta)^p}=\f{I_p(q)-I_{p-1}(q)}{q},\qquad
  \f{1}{2\pi}\int_0^{2\pi}\f{\sin\theta\,\rmd\theta}{(1-q\cos\theta)^p}=0.
\end{equation}
Thus, for $\gamma>1$,
\begin{equation}
  F^\mathrm{(2D)}=\f{1}{\gamma-1}\left[\f{\sqrt{1-e^2}}{1-e(e+ae_a)}\right]^{\gamma-1}[(1-x)I_{\gamma-1}(q)+xI_{\gamma-2}(q)],
\end{equation}
with
\begin{equation}
  x=\f{e\cos\alpha}{q}=\f{eae_a}{q^2[1-e(e+ae_a)]}=\f{eae_a[1-e(e+ae_a)]}{(ae_a)^2+(1-e^2)(ae\varpi_a)^2}.
\end{equation}
When $\gamma=2$ we have a result in terms of elementary functions:
\begin{equation}
  F^\mathrm{(2D)}=\f{\sqrt{1-e^2}}{1-e(e+ae_a)}\left(\f{1-x}{\sqrt{1-q^2}}+x\right).
\end{equation}
In the untwisted case this reduces to
\begin{equation}
  F^\mathrm{(2D)}=\f{1}{(f-e)}\left[\f{(1+e^2)f-2e}{\sqrt{1-f^2}}+e\sqrt{1-e^2}\right],
\end{equation}
where $f=e+ae_a$.

For the isothermal case $\gamma=1$ we use instead
\begin{equation}
  \f{1}{2\pi}\int_0^{2\pi}\ln(1-q\cos\theta)\,\rmd\theta=\ln\left(\f{\half q^2}{1-\sqrt{1-q^2}}\right),\qquad
  \f{1}{2\pi}\int_0^{2\pi}\cos\theta\,\ln(1-q\cos\theta)\,\rmd\theta=-\left(\f{1-\sqrt{1-q^2}}{q}\right),
\end{equation}
leading to
\begin{equation}
  F^\mathrm{(2D)}=\ln\left[\f{2\sqrt{1-e^2}\left(1-\sqrt{1-q^2}\right)}{q^2[1-e(e+ae_a)]}\right]-\f{eae_a\left(1-\sqrt{1-q^2}\right)}{q^2[1-e(e+ae_a)]}.
\end{equation}
In the untwisted case this reduces to
\begin{equation}
  F^\mathrm{(2D)}=\ln\left[\f{2\sqrt{1-e^2}\left(1-ef-\sqrt{1-e^2}\sqrt{1-f^2}\right)}{(f-e)^2}\right]-\left(\f{e}{f-e}\right)\left(1-ef-\sqrt{1-e^2}\sqrt{1-f^2}\right).
\label{f_gamma=1_untwisted}
\end{equation}

The apparent singularities at $e=f$ in the untwisted expressions can
be removed by using the trigonometric parametrization $e=\sin2\alpha$,
$f=\sin2\beta$, $\sqrt{1-e^2}=\cos2\alpha$, $\sqrt{1-f^2}=\cos2\beta$.
This leads to
\begin{equation}
  F^\mathrm{(2D)}=\f{1}{2}\sec2\beta\sec(\alpha+\beta)[3\cos(\alpha+\beta)-\cos(3\alpha-\beta)]
\end{equation}
for $\gamma=2$ and
\begin{equation}
  F^\mathrm{(2D)}=\ln\cos2\alpha-2\ln\cos(\alpha+\beta)+\sin2\alpha\sin(\alpha-\beta)\sec(\alpha+\beta)
\end{equation}
for $\gamma=1$.

\bsp	
\label{lastpage}
\end{document}